\documentclass[12pt,preprint]{aastex}
\shortauthors{\"Oberg et al.}

\begin{document}

\title{The Spatial Distribution of Organics toward the High-Mass YSO NGC 7538 IRS9}

\author{Karin I. \"Oberg}
\affil{Departments of Chemistry and Astronomy, University of Virginia, Charlottesville, VA
22904, USA}
\email{koberg@cfa.harvard.edu}
\author{Mavis Dufie Boamah}
\affil{Wellesley College, 106 Central Street, Wellesley, MA 02481, USA}
\author{Edith C. Fayolle}
\affil{Leiden Observatory, Leiden University, P.O. Box 9513, 2300 RA Leiden, The Netherlands}
\author{Robin T. Garrod}
\affil{Center for Radiophysics and Space Research, Cornell University, Ithaca, NY 14853-6801, USA}
\author{Claudia Cyganowski}
\affil{Harvard-Smithsonian Center for Astrophysics, 60 Garden Street, 
Cambridge, MA 02138, USA}
\author{Floris van der Tak}
\affil{Kapteyn Astronomical Institute, University of Groningen, The Netherlands / SRON Netherlands Institute for Space Research, Landleven 12, 9747 AD, Groningen, The Netherlands}

\begin{abstract}

Based on astrochemical theory, the complex molecular composition around high-mass YSOs should evolve from the outer envelope in toward the central hot region as a sequence of temperature dependent chemical pathways are activated in ices and in the gas-phase. The resulting complex molecules have been broadly classified into three generations dependent on the temperature ($<$25, $>$25, and $>$100~K) required for formation.  
We combine IRAM 30m and Submillimeter Array observations to explore the spatial distribution of organic molecules around the high-mass young stellar object NGC 7538 IRS9, whose weak complex molecule emission previously escaped detection, quantifying the emission and abundance profiles of key organic molecules as a function of distance from the central protostar. We find that emission from N-bearing organics and saturated O-bearing organics present large increases in emission around 8000~AU and R$<$3000~AU, while O-bearing molecules and hydrocarbons  do not. The increase in flux from some complex molecules in the envelope, around 8000~AU or 25~K, is consistent with recent model predictions of an onset of complex ice chemistry at 20--30~K. The emission increase for some molecules at R$<$3000~AU suggests the presence of a weak hot core, where thermal ice evaporation and hot gas-phase chemistry drives the chemistry. Complex organics thus form at all radii and temperatures around this protostar, but the composition changes dramatically as the temperature increases, which is used to constrain the chemical generation(s) to which different classes of molecule belong. 
 
\end{abstract}

\keywords{astrochemistry --- 
circumstellar matter --- 
molecular processes ---
massive star formation ---
individual object: NGC7538-IRS9}

\section{Introduction}

The early phases of high-mass ($M_{\rm zams}>8M_\odot$) star formation take place in highly obscured regions, where the central young stellar object (YSO) is embedded in a large envelope of cold, luke-warm, and hot gas and dust. Because the YSO is deeply obscured, observations of trace species, i.e. dust and gas-phase molecules other than H$_2$, often provide the best and sometimes only constraints on the early evolution of these stars \citep{vanDishoeck98}. The utility of such molecular probes depend, however, strongly on how well their chemistry is understood, and especially how their formation and destruction efficiencies depend on the environment.

Complex organic molecules are common probes of the 'hot core' stage of massive young stellar objects (MYSOs), which is characterized by intense emission of different organic molecules from a dense and hot region close to the protostar \citep{Blake87,Helmich97,Schilke01}. Complex organic molecules are proposed to form through ice chemistry in the colder parts of the protostellar envelope and then evaporate as material flows toward the protostar. While organic ices evaporate over a range of temperatures, most of the ice is expected to be released at $\sim$100~K, forming the observed hot core \citep{Viti04,Nomura04,Garrod08}. The molecular composition may then further evolve through gas-phase chemistry \citep[e.g.][]{Charnley95}.

 \citet{Herbst09} categorized observed complex molecules into zeroth, first and second generation species, where zeroth generation molecules form through cold (10~K) hydrogenation grain-surface/ice reactions (e.g. CH$_3$OH from CO, \citep{Tielens82}) or cold gas-phase chemistry, first generation from photodissociation of the zeroth generation ices followed by radical diffusion and recombination in the icy grain mantles \citep{Garrod08}, and second generation from hot gas-phase chemistry following complete evaporation of zeroth and first generation species into the gas-phase. With the exception of a few molecules, it is however unclear which complex molecule falls into which category, and some, e.g. CH$_3$OCH$_3$, are probably produced both through first and second generation chemistry \citep{Bisschop07,Garrod08,Herbst09}.

In this scenario, the chemical composition in an isolated MYSO, where the envelope is mainly heated by the central protostar, should change radically as a function of radius. In the outer envelope (here defined as the radii beyond which T$<$25~K) radical diffusion in ices is slow and only zeroth generation molecules should be present. In the inner envelope, spanning the radii where 25~K$<$T$<$100~K, complex ice chemistry is expected to be efficient, and through non-thermal desorption \citep[e.g.][]{Garrod07,Oberg10a,Oberg10b} some of the products should be released into the gas-phase, resulting in a mixture of zeroth and first generation molecules. We define the hot core region as the radii where T$>$100~K, where all ices evaporate on short timescales and high density gas-phase chemistry may produce second generation products. The size scales of these different regions will vary between different objects, dependent on the luminosity of the central protostar and the envelope density profile \citep{vanderTak00}.

Observations of chemical differentiation within a single MYSO has a great potential to test this hypothesis on how  complex molecule form at different temperatures, and to constrain which generation specific molecules belong; cf. \citet{JimenezSerra12} where observations of chemical differentiation within a hot core was used to test molecular destruction models.  Observations of envelope chemistry are challenging, however, toward typical complex chemistry sources, where the hot core emission dominates, e.g. NGC7538 IRS 1. In these sources, the emission from the hot core easily drowns out any envelope emission from complex organics. Based on the results from a recent small survey of MYSOs (Fayolle et al. in prep.), we instead target a MYSO, NGC7538 IRS 9, whose weak complex molecule line emission can only in part be attributed to a central unresolved core. 

The cloud region NGC7538, located at 2.65[0.12] kpc \citep{Moscadelli09}, harbors  several massive YSOs, which display a large range in chemical characteristics, including the hot core of NGC 7538 IRS1 -- the effect of this source on the chemistry in nearby, less bright sources is unknown and is evaluated based on molecular emission structures in \S\ref{sec:res}. NGC7538 IRS9 is estimated to contain about 1000~M$_\odot$ (virial mass within $7\times10^4$~AU) and has a bolometric luminosity of $6\times10^4$L$_{\odot}$ \citep{vanderTak00}. It has at least three bipolar molecular outflows \citep{Sandell05}, which indicates the presence of multiple YSOs within larger structure. NGC7538 IRS9 is also associated with water, Class I methanol, and OH masers, and possibly with Class II methanol masers \citep{Kameya90,Sandell05,Sugiyama08, Pestalozzi06}. The combination of multiplicity and outflows will complicate the interpretation of the chemistry, but the overall chemical evolutionary state at any given point may still be dominated by the distance to the central, most luminous protostar. The chemistry of NGC 7538 IRS9 is relatively unknown, but it has strong ice absorption bands, indicating precursors of complex molecules are common in the outermost envelope. No hot CH$_3$OH was detected when observed with a single dish \citep{vanderTak00b}, but infrared absorption and emission lines from a hot (T$>$100~K) component have been detected \citep{Boogert04c,Barentine12}. 

In this paper we use a combination of single dish and interferometric spectrally resolved observations to explore how the chemistry depends on the distance from the center of the protostar, and interpret the results in light of a state-of-the-art astrochemical model. We focus on the major trends, with more detailed modeling to follow in future papers. In \S\ref{sec:obs} we present 1mm, spectrally resolved IRAM 30m and Submillimeter Array (SMA) observations toward NGC 738 IRS9, and describe how the combined data set have been calibrated to be self-consistent. The extracted spectra, images and qualitative and quantitative results on the distribution of organic molecules toward this MYSO are presented in \S\ref{sec:res}. The implications of the results in terms of the chemical evolution toward NGC 7538 IRS9, and predictions of a new model by Garrod (ApJ, accepted), mapped onto the temperature structure of IRS9 are discussed in \S\ref{sec:disc}. \S\ref{sec:conc} presents some concluding remarks.

\section{Observations \label{sec:obs}}

\subsection{IRAM 30m}

NGC 7538 IRS9 was observed with the IRAM 30m Telescope February 19--20, 2012 using the EMIR 230 GHz receiver and the new FTS backend. The two sidebands cover 223--231~GHz and 239--247~GHz at a spectral resolution of $\sim$0.2~km~s$^{-1}$ and with a sideband rejection of -15dB \citep{Carter12}. The pointing position was Ra=23:14:01:60, Dec=61:27:20.4 and pointing was checked every 1--2~h and found to be accurate within 2--3$\arcsec$. Focus was checked every 4~h, and generally remained stable through most of the observations, i.e. corrections on the order of 0.2--0.4 was common, but a correction of 0.7 was required once. Observations were acquired using both the position switching and wobbler switching modes. The position switching mode was attempted because of possible extended emission. However this setting resulted in instabilities and we therefore switched to the wobbler mode. Comparison of the two spectra  reveals no significant absorption in the wobbler off-position, hence we only use the higher-quality wobbler spectra in this paper. The comparison did however reveal some calibration inconsistencies between the spectra acquired in the two modes (see Appendices), but this does not affect the results since we used the calibrated SMA spectra for absolute flux calibration (see below). The total integration time in the wobbler mode was $\sim$4~h under excellent weather conditions, resulting in a T$_{\rm a}$ rms of 8~mK. 

The spectra were reduced using CLASS. A global baseline was fitted to each 4~GHz spectral chunk using 4--7 windows. The individual scans were baseline subtracted and averaged. To convert from antenna temperature, T$_{\rm a}$, to main beam temperature, T$_{\rm mb}$, forward efficiencies and beam efficiencies of 0.92 and 0.60 were applied for the lower spectral chunks and 0.90 and 0.56 for the upper chunks. In this paper we only discuss the upper inner and lower inner spectral chunks (227.05--231.10~GHz and 239.00--243.05~GHz), which overlap almost completely with SMA observations toward the same source. 

\subsection{Submillimeter Array}

NGC 7538 IRS 9 was observed with the Submillimeter Array\footnote{The Submillimeter Array is a joint project between the Smithsonian Astrophysical Observator and the Academia Sinica Institue of Astronomy and Astrophysics. It is funded by the Smithsonian Institute and the Aademia Sinica.} (SMA) August 15th (compact configuration) and October 15th (extended configuration), 2011 at good to excellent weather ($\tau_{225\,\rm GHz}$ = 0.06--0.15). The phase center of the observations were Ra=23:14:01.68 Dec=61:27:19.1, i.e. within 1.5$\arcsec$ of the IRAM 30m pointing position.

The combined range of baselines was 16--226~m. The SMA correlator was set-up to obtain a spectral resolution of $\sim$1 km s$^{-1}$ using 128 channels for each of the 46 chunks covering 227-231~GHz in the lower sideband and 239--243~GHz in the upper sideband. The observing loops used J2202+422 and J0102+584 as gain
calibrators. Flux calibration was done using observations of Uranus and Callisto. The derived
flux of J2202+422 was 4.04 Jy (August 15th), and of J0192+584 2.46~Jy (August 15th) and 1.81 (October 15th). The bandpass response was calibrated using observations of 3C84.
Routine calibration tasks were performed using the MIR software package, and
imaging and deconvolution were accomplished in the MIRIAD software package. 

The continuum was subtracted separately for the upper and lower sideband for each observational data set in MIR, using line-free channels. The continuum-subtracted compact and extended data were then combined with MIRIAD using natural weighting. The resulting synthesized beam is $2.3\arcsec\times2.0\arcsec$. The primary beam of the SMA at these wavelengths is $\sim$50$\arcsec$. Considering the baseline coverage, all emission at scales larger than 18$\arcsec$ is completely filtered out, but followed the calculation of \citet{Wilner94}, up angular structures smaller than 7$\arcsec$ is required to filter out less than 50\% of the emission.

\subsection{IRAM 30m Flux Correction}

NGC 7538 IRS 9 spectra were acquired at the IRAM 30m as a part of a larger observational project that also involved NGC 7538 IRS1. Initial inspection of the NGC 7538 IRS 1 IRAM 30m spectra (converted to main beam temperature using the recommended conversion factors at http://www.iram.es/ IRAMES/mainWiki/Iram30mEfficiencies) revealed a factor of two lower line intensities compared to two previous studies, for all overlapping transitions,  \citep{vanderTak00,Bisschop07}. This fact together with the previously noted intensity discrepancy between spectra acquired in wobbler and position-switch mode (and the recent commissioning of the new FTS at the time of observations) motivated us to recalibrate the IRAM 30m spectra with respect to the carefully flux-calibrated SMA spectra. Because of spatial filtering of extended emission in the interferometric data we only used high-excitation lines of molecules expected to be exclusively present close to the protostar, i.e. CH$_3$CN and vibrationally excited CH$_3$OH (only detected in NGC 7538 IRS1). To match the flux of these lines in the IRAM and SMA spectra requires the IRAM spectra to be multiplied by 1.8[0.2] in addition to the recommended T$_{\rm a}$-to-Flux conversion factor of 8.1. The origin of this mismatch is not completely understood, but a fraction of it is probably due to the declination dependent antenna gain, which was not accounted for explicitly. Slightly out of focus observations, resulting in a larger beam may also contribute. The absolute fluxes obtained at the SMA also has a typical uncertainty of at least 10\%, but uncertainties up to 20\% cannot be excluded. Thus, while the IRAM 30m and SMA observations can be calibrated to one another, enabling accurate determinations of relative abundances, the absolute column densities are estimated to be accurate only within a factor of two.

\section{Results \label{sec:res}}

\subsection{Molecular Line Identifications and Selections}

\begin{figure*}[htp]
\epsscale{1.0}
\plotone{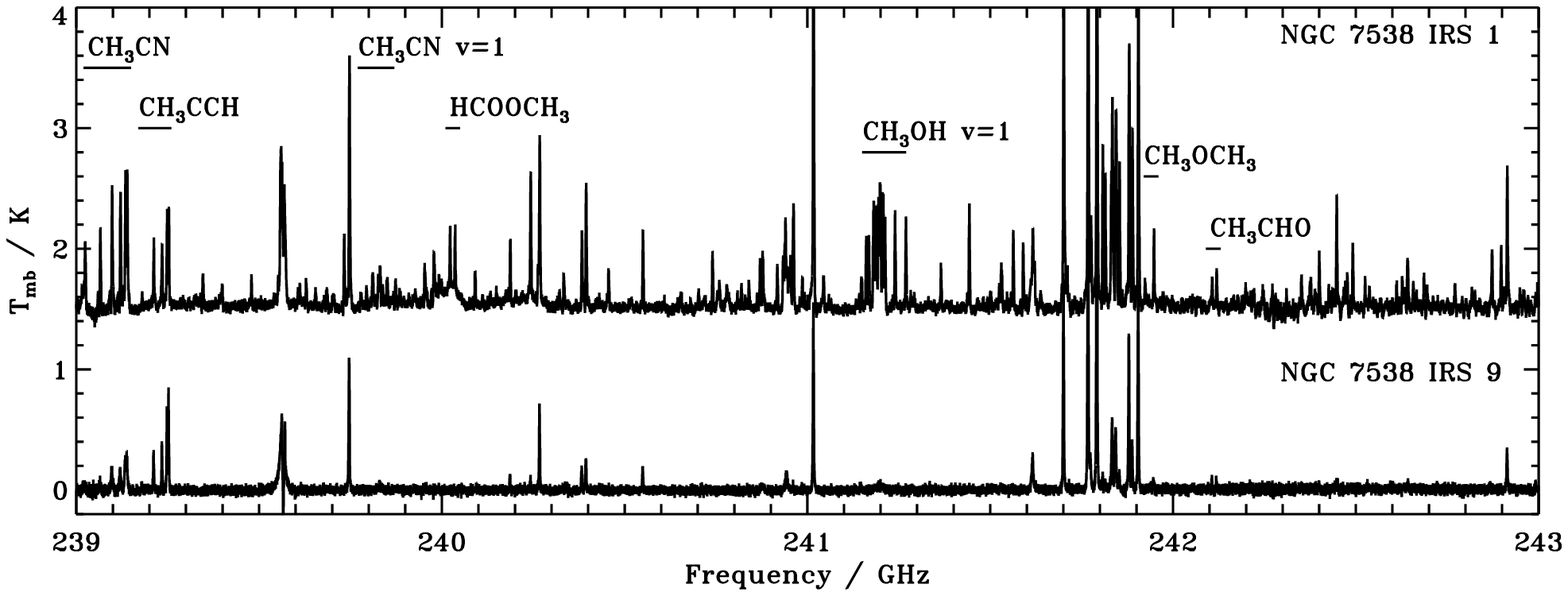}
\caption{239--243~GHz spectra of NGC 7538 IRS 1 and IRS 9, displaying the difference in line emission from organic molecules in the two sources. \label{fig:irs1}}
\end{figure*}

Figure \ref{fig:irs1} shows the IRAM 30m spectra at 139--243~GHz toward NGC 7538 IRS 9, and how it compares with the same spectra toward the hot-core source NGC 7538 IRS1 (observed under the same conditions). While there are a large number of line coincidences between the two sources, NGC7538 IRS9 displays both weaker lines and a lower line density. 
The main limitation of the IRS9 dataset is therefore poor S/N rather than line overlaps, and identifications of detected lines are quite straightforward compared to hot cores.

Lines were identified in the IRAM 30m spectra by first checking for frequency coincidences with the line compilation Splatalogue, drawing upon the CDMS and JPL spectral databases \citep{Pickett98,Muller01}. We assumed the literature V$_{\rm lsr}=-57$ km/s \citep{vanderTak00}, and this value was readily confirmed with our own data. For the initial analysis, only lines with Einstein coefficients $A_{\rm ij}>-5$ and upper energy levels $E_{\rm up}<300$~K were considered, since low temperatures and low column densities of organic molecules were expected \citep{vanderTak00b}. Most lines could be assigned using these constraints. The remaining unassigned lines were checked for matches in the databases without any energy level and Einstein coefficient constraints, but this did not result in any additional identifications. In addition we identified two ghosts from CO and H$_2$CO, consistent with the sideband rejection of -15.

Most preliminary line identifications implied the presence of other lines in the observed range, i.e. the catalogue contained other transitions of the same species with comparable  $A_{\rm ij}$ and $E_{\rm up}$ values. The final line list (Appendix A) only includes species that presented all expected lines within the observed spectral region. From this line list, we selected lines and line ladders belonging to nine different organic molecules to analyze in greater detail. These molecules were all detected in at least three different lines in the IRAM~30m data, with the exception of $c$-C$_3$H$_2$, which only presents one strong transition within the surveyed frequency range (at 227.169~GHz) and HC$_3$N, which only has two transitions (at 227.419 and 245.606~GHz). Notable non-detections include HCOOCH$_3$, CH$_3$CH$_2$OH and C$_2$H$_5$CN. Most, but not all, lines detected in the IRAM~30m spectra could also be identified in the SMA spectra -- the exceptions are very weak lines (the S/N is better in the IRAM~30m spectra) and a few stronger lines that were inferred to be spatially filtered out.

\subsection{SMA Molecular Images}

\begin{figure*}[htp]
\epsscale{1.5}
	\includegraphics[width=\linewidth]{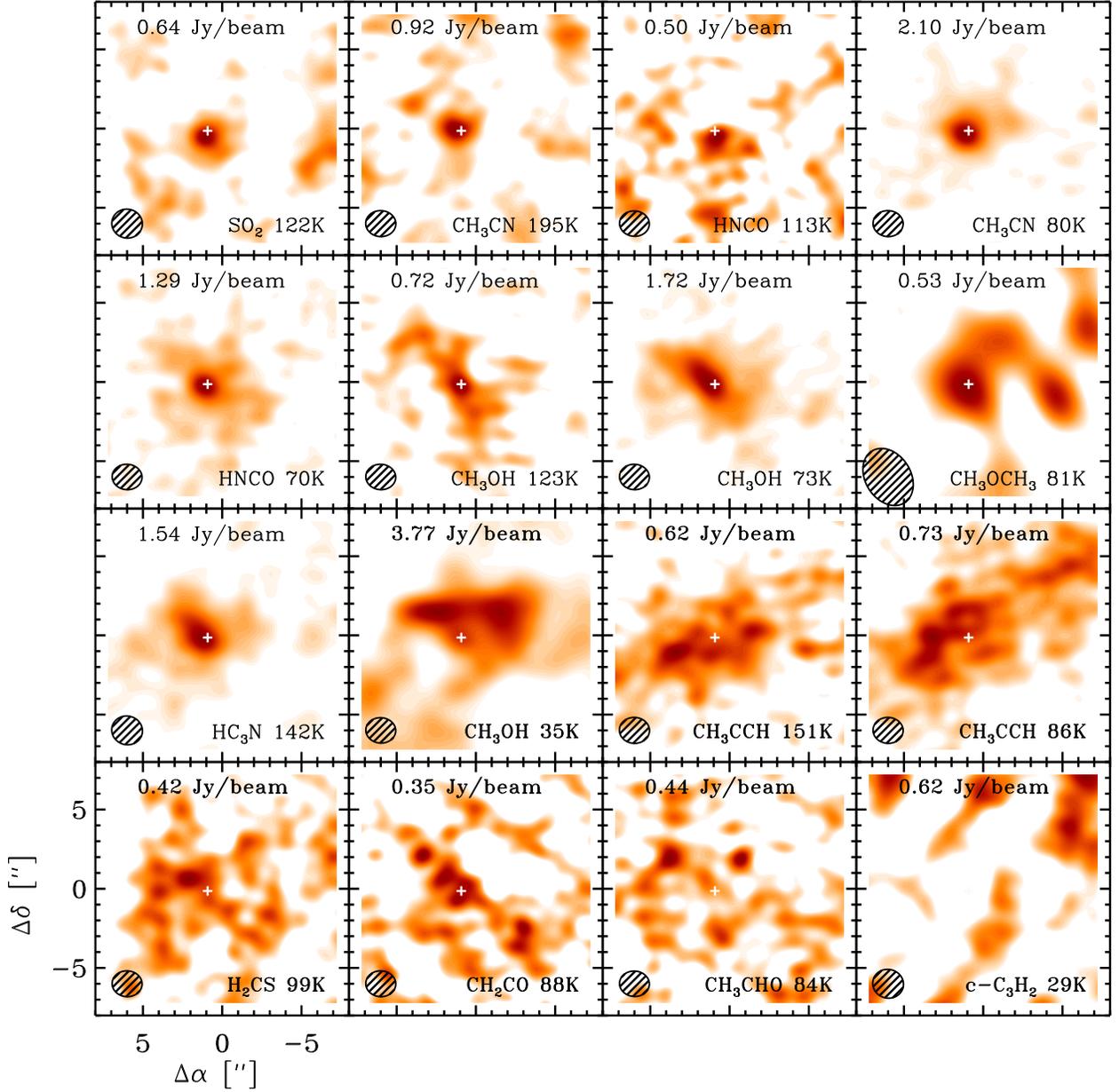}
\caption{Submillimeter integrated flux maps of key transitions ordered from centrally condensed to diffuse. The synthesized beam is shown in the lower left corner (for CH$_3$OCH$_3$ the longest baseline were not included to better display this tentative detection). The upper energy level of each line is listed next to the molecular formula and the peak integrated flux is listed in  Jy km/s beam$^{-1}$ at the top of each panel. The black contours mark 50\% of the emission peaks. All molecules in the top three rows are detected, with the exception for the tentative CH$_3$OCH$_3$ detection, while no detections are claimed for the four transitions in the bottom row. \label{fig:maps}}
\end{figure*}

Figure \ref{fig:maps} displays the spatially resolved emission, integrated over each line, for 16 transitions belonging to 11 different molecules, ordered from centrally concentrated to diffuse (based on visual inspection of the images). Only SO$_2$ and high-energy (E$_{\rm u}>$100~K) transitions of N-bearing saturated organic molecules (i.e. not the unsaturated carbon chain HC$_3$N) present completely unresolved emission. Lower-energy transitions of N-bearing organics, as well as CH$_3$OH lines with excitation levels of 80--120~K comprise a core and an extended component. CH$_3$OCH$_3$ seems to have a compact component, but the S/N is very low. In contrast CH$_3$CCH and H$_2$CS are dominated by extended emission, and CH$_3$CHO, CH$_2$CO and $c$-C$_3$H$_2$ are undetected in the SMA images, indicative of a smooth extended distribution that is filtered out -- these lines are all clearly detected in the IRAM 30m spectra.

These images clearly reveal that even traditional hot core molecules, such as CH$_3$CN, have extended components when observing low-energy transitions. The extended component is not nearly as bright as for some O-bearing complex molecules, however; when comparing emission lines from different molecules with similar excitation properties, there are qualitative differences in the hydrocarbon, O- and N-bearing complex chemistry, similarly to what has been previously noted in \citet{Caselli93,Wyrowski99}. These differences cannot be attributed to differences in optical depth -- as discussed below all lines seem to be optically thin -- or to excitation conditions.

To provide an initial estimate of the symmetry of the molecular distribution, a 2D Gaussian was fit to the visibilities of each transition in Fig. \ref{fig:maps}. Meaningful parameters were derived for the CH$_3$OH, 80~K CH$_3$CN, CH$_3$CCH, 70~K HNCO, H$_2$CS and HC$_3$CN lines. The remaining transitions presented either insufficient S/N or completely unresolved emission. The fitted Gaussians have minor/major axis ratios of 0.6--1.0 with an average of 0.7. The images are thus relatively axisymmetric, justifying the next step in the analysis. We also checked for systematic offsets in emission along previously discovered outflow axes, and toward IRS 1 (NW), but as Figure \ref{fig:maps} shows, neither kind of asymmetry is present at an appreciable level. 

\subsection{SMA Image Analysis}

\begin{figure}[htp]
\epsscale{1.0}
\plotone{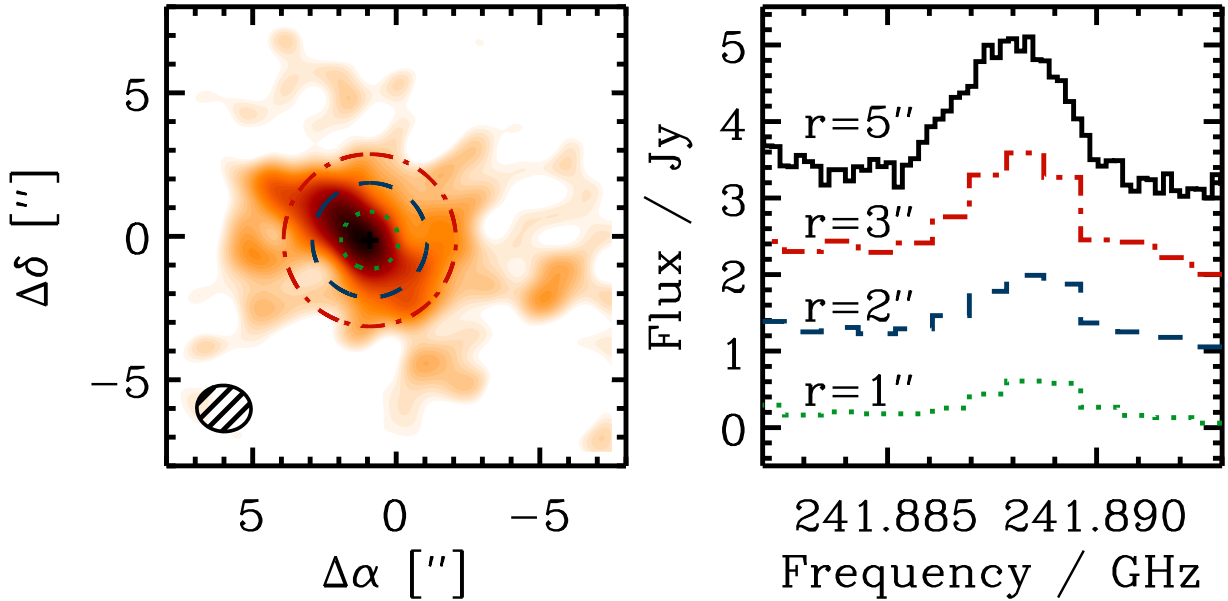}
\caption{{\it Right:} The image of a CH$_3$OH line, plotted together with the three masks used to extract spectra from the SMA data. {\it Left:} The resulting spectra plotted together with the IRAM 30m spectra which encompass a beam with a radius of 5$\arcsec$. \label{fig:mask}}
\end{figure}

Spectra are extracted from the spatially and spectrally resolved SMA data cubes by applying three circular masks during the data reduction, centered on the continuum peak in the SMA dataset, with radii of 1, 2, and 3$\arcsec$, respectively (Fig. \ref{fig:mask}). The resulting spectra will contain increasing line fluxes as more protostellar material is encompassed; steep increases imply that the line emission is distributed throughout the envelope, while a shallow increase imply that the emission is concentrated toward the central region. The mask sizes are selected such that the smallest one coincides with the beam size and the largest one with a physical radius of $\sim$8000~AU, where the kinetic temperature should drop below 25~K according to the temperature and density model developed for this object \citep{vanderTak00}. At lower temperatures, significant complex molecule formation is not expected to occur \citep{Garrod08}. Using the extracted line fluxes from the SMA data and the calibrated IRAM 30m line fluxes, the fluxes in annuli between 0 and 5$\arcsec$ can then be determined. These annular fluxes form the foundation for our analysis and data-model comparison.

This strategy to quantify the spatially resolved chemistry around a protostar depends on an assumption that spectra extracted from the same radii, but at different azimuth can be combined. This is a reasonable assumption if the protostellar chemistry is regulated by the temperature set by the distance from the central protostar, as is often assumed in models \citep[e.g.][]{vanderTak00,Garrod08}. This simple picture will break down in sources where outflows instead dominate the observed chemistry, which is difficult to predict {\it a priori}, but a spherical symmetry seems to be a reasonable starting point toward NGC 7538 IRS9 based on the symmetry of the molecular line images in Fig \ref{fig:maps}.

\subsection{Radial Distributions of Line Emission}

\begin{figure}[htp]
\epsscale{0.7}
\plotone{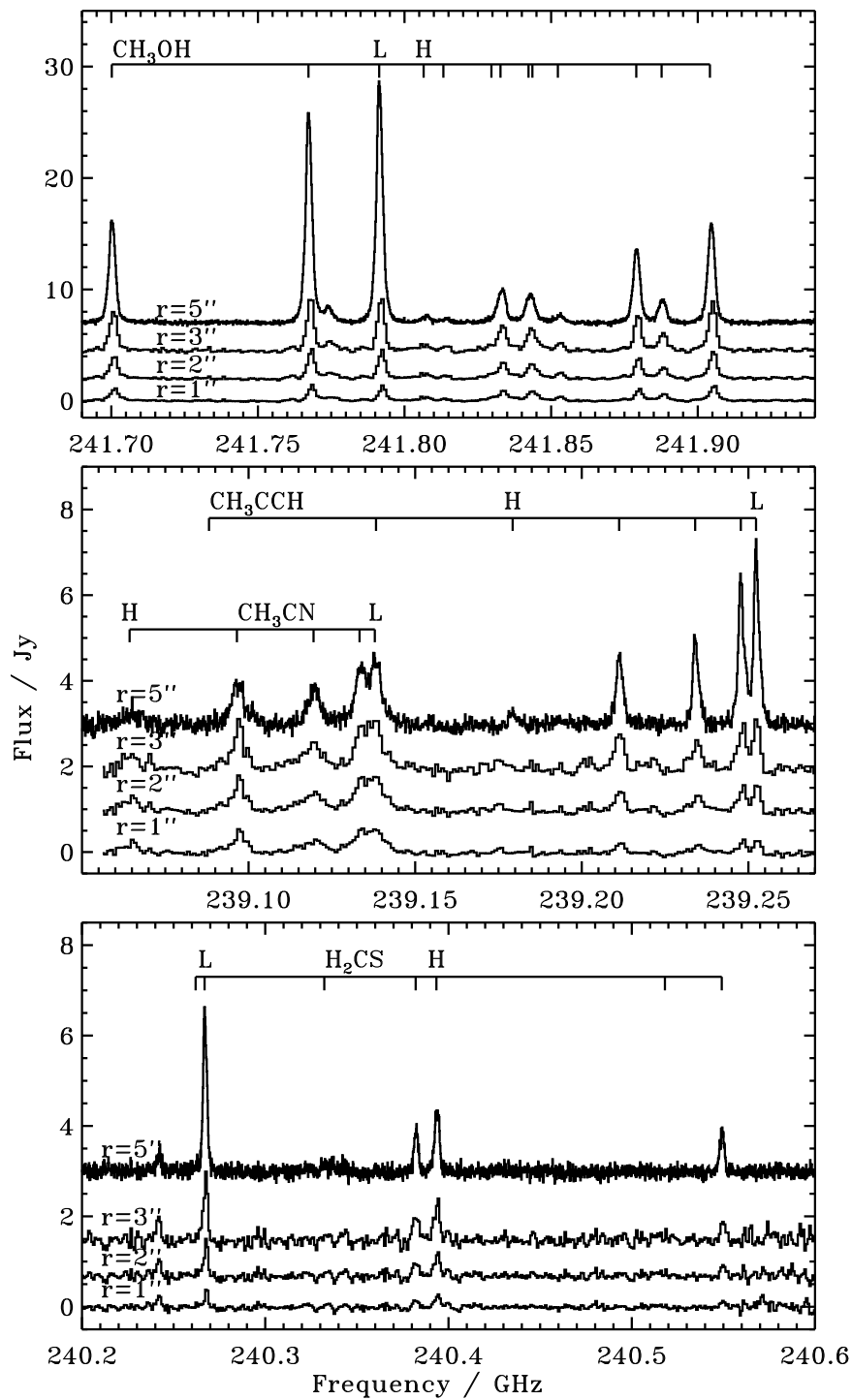}
\caption{Spectra of detected lines from the CH$_3$OH $5-4$, CH$_3$CN $13-12$, CH$_3$CCH $14-13$ and H$_2$CS $7-6$ ladders, extracted from the SMA data using masks centered on the continuum peak with radii 1--3$\arcsec$, and from the IRAM data, which has a beam size of $\sim$10$\arcsec$. The detected transitions with the highest and lowest energy levels are marked. \label{fig:ladders}}
\end{figure}

\begin{figure}[htp]
\epsscale{0.8}
\plotone{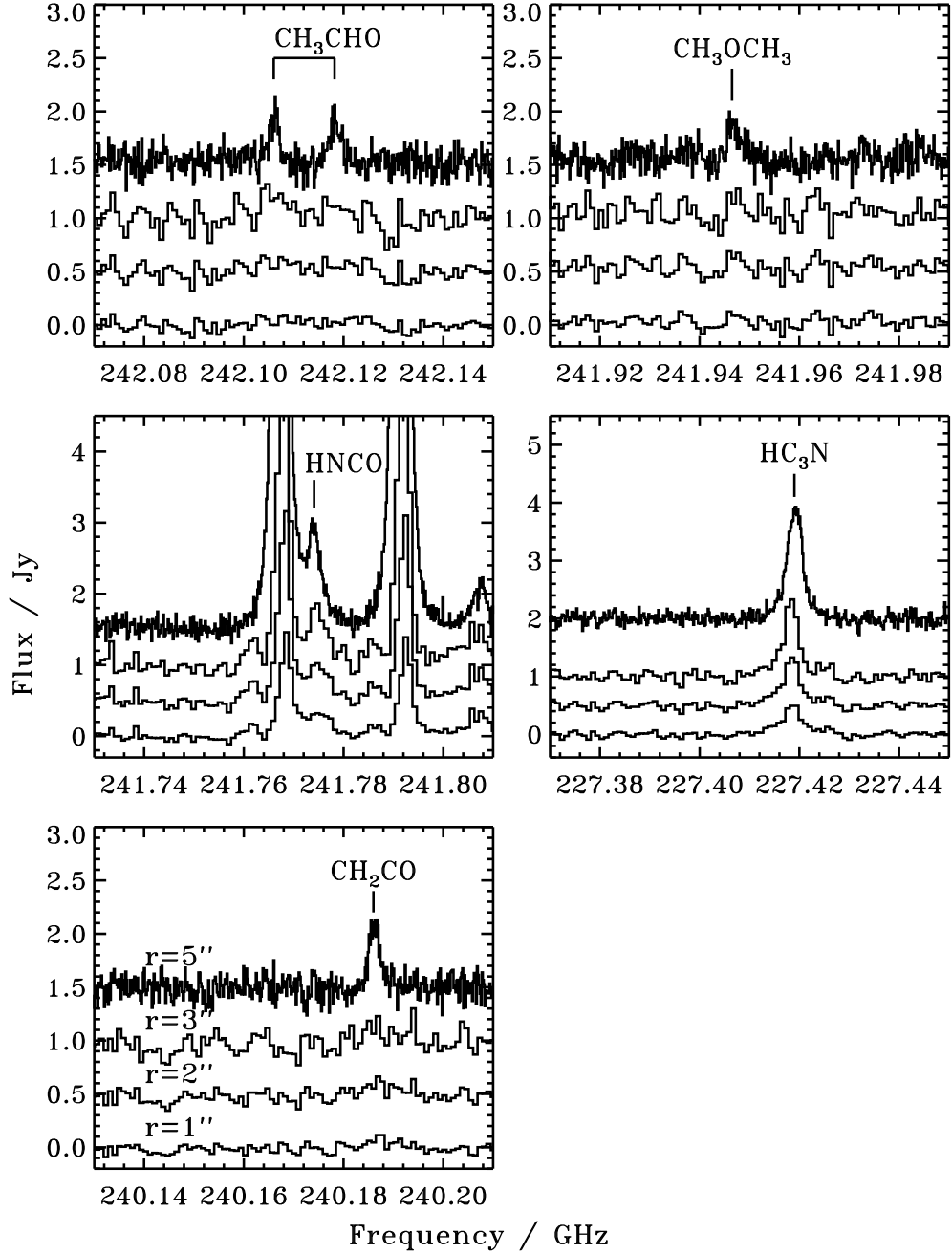}
\caption{Spectra of three complex organics, extracted from the SMA data using masks centered on the continuum peak with radii 1--3$\arcsec$ and the IRAM spectra with a beam size of $\sim$10$\arcsec$. CH$_3$CHO and CH$_2$CO are only detected with IRAM and the SMA upper limits are significant. In contrast there is a hint of a CH$_3$OCH$_3$ line in the SMA spectra, but the current rms level makes this a tentative detection. \label{fig:comp}}
\end{figure}

Figures \ref{fig:ladders} and \ref{fig:comp} show extracted spectra of CH$_3$OH, CH$_3$CN, CH$_3$CCH, H$_2$CS, CH$_3$CHO, CH$_3$OCH$_3$, HNCO, HC$_3$CN and CH$_2$CO using the three SMA masks (radius$=$1, 2 and 3$\arcsec$) and the IRAM 30m spectra (corresponding to a beam radius of 5$\arcsec$). Figure \ref{fig:ladders} clearly demonstrates that the relative fluxes of different transitions within each ladder changes when the spectra are extracted using different masks. In the IRAM 30m spectra the low-energy transitions generally dominate, while in the central core the flux is more equally partitioned among the low and high-energy transitions. This is consistent with a radial temperature profile, regulated by heating from a central source -- i.e closer to the core the temperature will be higher, resulting in more excited molecules.

The differences between N-bearing molecules, saturated and unsaturated O-bearing molecules, the hydrocarbon CH$_3$CCH, and the S-bearing molecule H$_2$CS are also apparent in the extracted spectra. Most  CH$_3$OH, CH$_3$CCH and H$_2$CS line fluxes decrease significantly when the emission region is decreased, i.e. only a fraction of the emission originate at small scales. Spectra of N-bearing molecules display much smaller differences between the spectra extracted with different masks, implying that a large fraction of the emission originates in the central 2$\arcsec$ core. Two molecules, CH$_3$CHO and CH$_2$CO, are undetected in the SMA spectra as expected from the images in Fig. \ref{fig:maps}. 

The integrated line fluxes in Figs. \ref{fig:ladders} and \ref{fig:comp} are listed in Table \ref{tab:flux}, except for a few lines that are too confused. Integrated line fluxes are obtained by first fitting Gaussians to the IRAM 30m spectra in velocity space using the IDL routine MPFIT. The Gaussians are fit to a small window around each listed peak position ($\pm$10--15~km s$^{-1}$) and include a local linear baseline fit. In crowded regions multiple Gaussians are fitted simultaneously. In most cases the SMA spectra are not of sufficient quality to do similar 5--10 Gaussian parameter fits and the peak position and FWHM extracted from the IRAM 30m fits to the same line(s) is therefore assumed. The resulting fits were inspected for each line. Upper limits in the SMA spectra are calculated using the measured rms in the vicinity of the line and the line width derived from the IRAM~30m spectral fit.

\begin{figure}[htp]
\epsscale{1.0}
\plotone{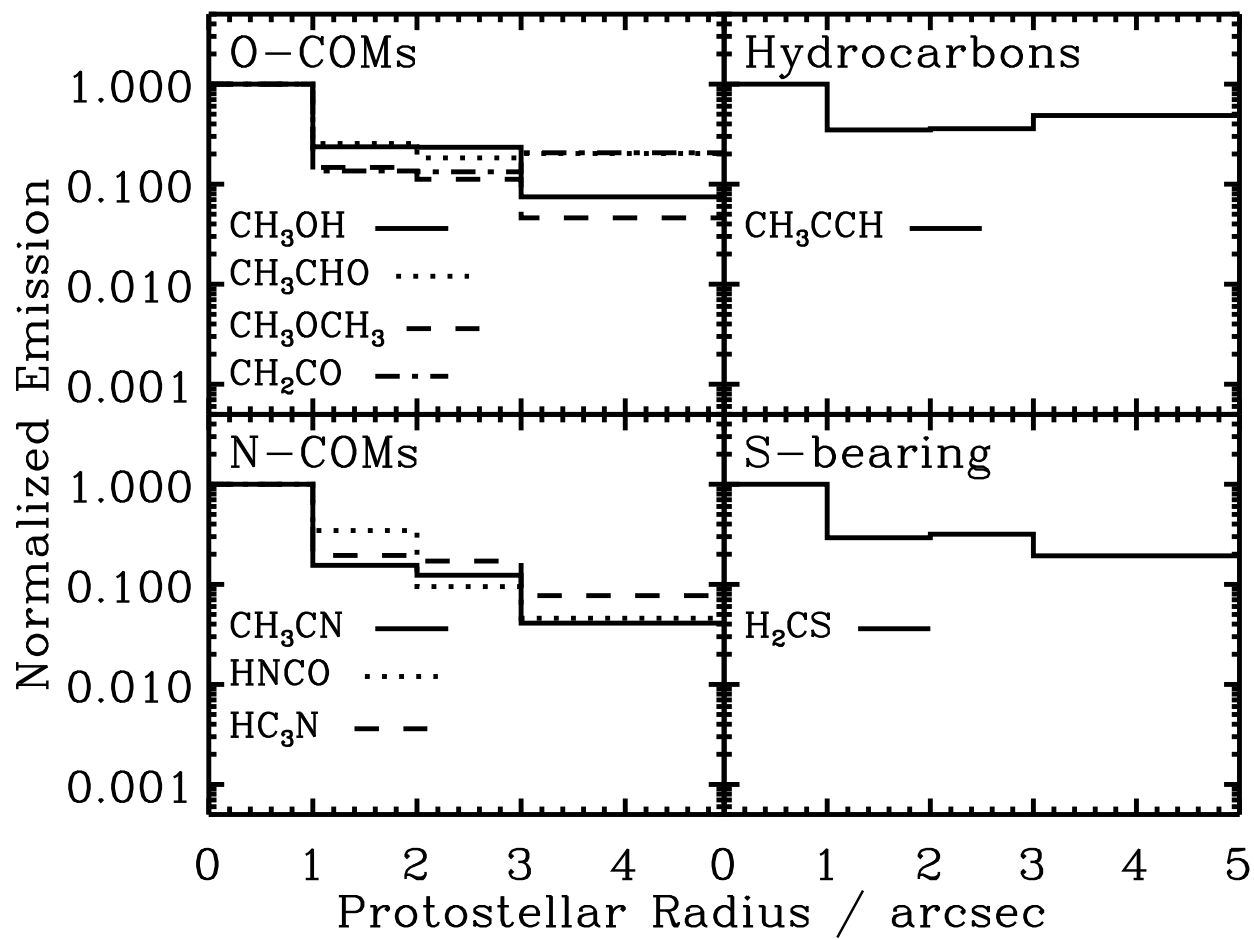}
\caption{Flux distributions normalized to the emission area and the central peak emission as a function of protostellar envelope radii for identified organic molecules. For CH$_3$OH, CH$_3$CCH, CH$_3$CN and H$_2$CS the transition with an excitation temperature closest to 80~K is shown. The CH$_3$CHO and CH$_2$CHO emission interior to 3$\arcsec$ and the CH$_3$OCH$_3$ emission interior to 2$\arcsec$ are upper limits. \label{fig:cum}}
\end{figure}

Based on Figs.  \ref{fig:ladders} and \ref{fig:comp}, the radial molecular emission distribution clearly depends on both species and excitation characteristics. To compare the emission patterns of different molecules in a meaningful way therefore requires lines with similar excitation levels. Fortunately all targeted molecules, except for HC$_3$N, have detected lines with upper energy levels of 70--100~K  (Table \ref{tab:flux}). Figure \ref{fig:cum} shows the integrated line fluxes from transitions with upper energy levels of $\sim$80~K (marked bold in Table \ref{tab:flux}) and the HC$_3$N line, in four annuli around the source center, normalized to the annulus emission area and to the flux at $r<1\arcsec$. It is important to keep in mind that because of spatial filtering, the flux at small scales should be interpreted as sharp flux enhancements on top of smooth protostellar envelope fluxes. 

Figure \ref{fig:cum} shows that seven of the investigated molecules display significant flux increases toward the center: CH$_3$OH, CH$_3$OCH$_3$, CH$_3$CN, HNCO, HC$_3$N, CH$_3$CCH and H$_2$CS. With the exception of CH$_3$CCH, the flux increase occurs in two steps, at $r\sim3\arcsec$ and at $r<1\arcsec$; CH$_3$CCH only displays an increase at the smallest scales. CH$_3$CHO and CH$_2$CO show no flux increase toward the protostellar center, but the upper limits at the smallest scales are too large to exclude an increase in flux at $r<1\arcsec$. Among the molecules that do display compact emission, the core to outer envelope ($r>3\arcsec$) flux ratio varies between a factor of 20 (CH$_3$CN and HNCO), and a factor of 2 (CH$_3$CCH), indicative of a changing organic composition throughout the protostellar envelope and core region. Trends in emission profiles cannot be directly translated into column densities, however, since they also depend on changes in excitation conditions. It is important not to over-interpret this result beyond that 1) some molecules are much more centrally peaked than others, 2) most species present some emission in the cold and extended envelope, and 3) there are two radii at which the chemical composition changes dramatically, only one of which can be attributed to a central hot core.

\subsection{Excitation Temperatures and Rotational Diagrams}

Four molecules, CH$_3$OH, CH$_3$CCH, CH$_3$CN and H$_2$CS, present sufficient number of transitions to derive excitation temperatures and column densities at different radii around NGC 7538 IRS9 using the rotation diagram method as formulated in \citet{Goldsmith99}, assuming a single excitation temperature for each molecule in each annulus and optically thin lines.  Both these assumptions can be evaluated by inspecting the rotational diagrams and the derived column densities of each molecule. No isotopologues were detected which implies that even the strong CH$_3$OH transitions are at most marginally optically thick (if the CH$_3$OH lines are optically thin and $^{12}$CH$_3$OH/$^{13}$CH$_3$OH$\sim$70, the strongest $^{13}$CH$_3$OH line should have a flux of 0.3 Jy km s$^{-1}$, just below the current detection limit). 

Table \ref{tab:cd} displays the calculated excitation temperatures and column densities. Because of spatial filtering, the column densities derived at small scales should be interpreted as excess column on top of the smooth background column density, i.e. there must be an increase in column between the smooth envelope and the 3--2$\arcsec$ annulus, otherwise no emission would be detected. This absolute column density increase between the smooth envelope and the smaller structures is not quantified. Instead we only investigate how the relative abundances changes between different protostellar regions.

\begin{figure}[htp]
\epsscale{1.0}
\plotone{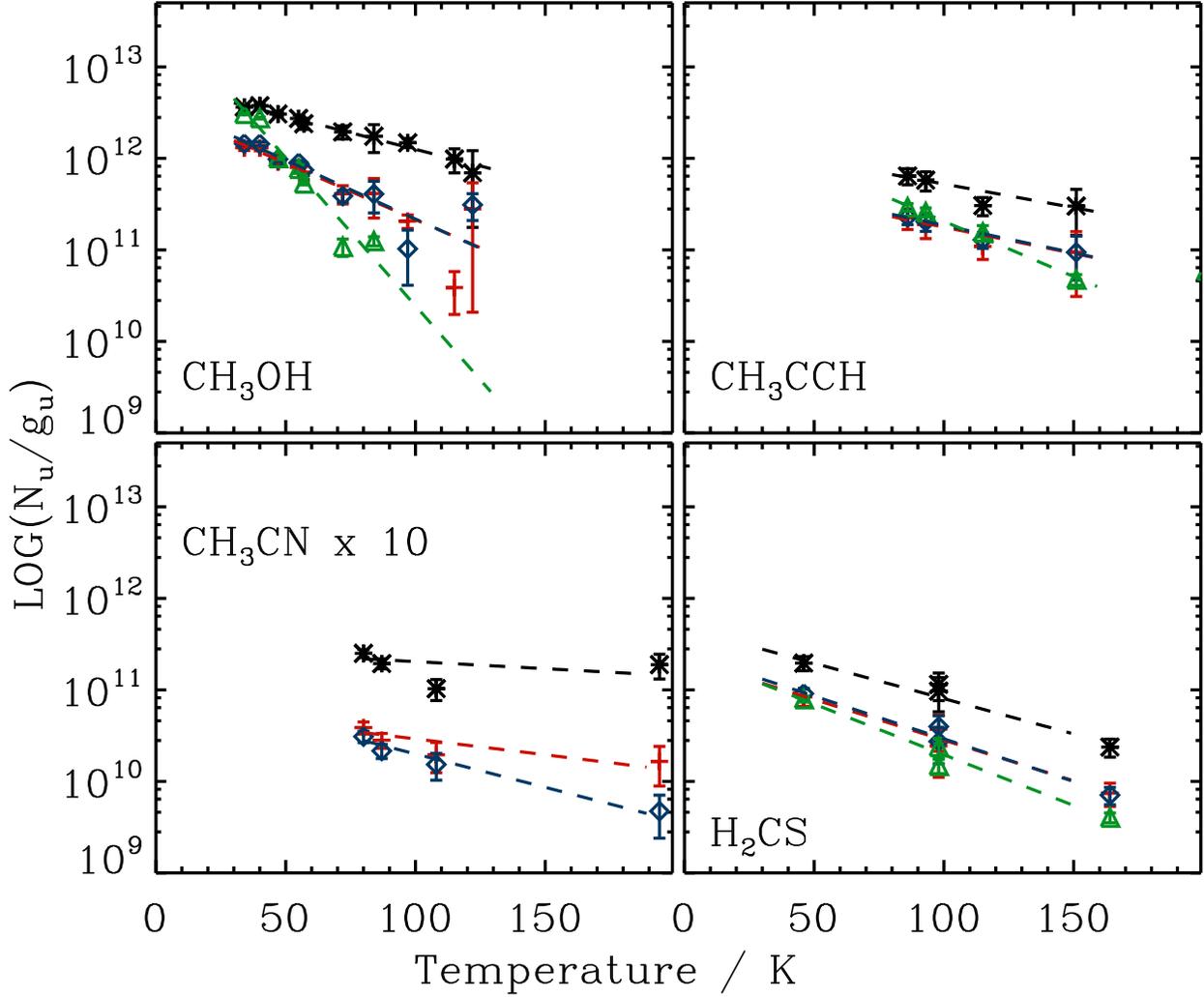}
\caption{CH$_3$OH, CH$_3$CCH, CH$_3$CN and H$_2$CS rotational diagrams based on the flux collected with the 1$\arcsec$ mask (black stars) and within 1--2$\arcsec$ (red crosses), 2--3$\arcsec$ (blue diamonds) and 3--5$\arcsec$ (green triangles) annuli. \label{fig:rot}}
\end{figure}

\begin{figure}[htp]
\epsscale{1.0}
\plotone{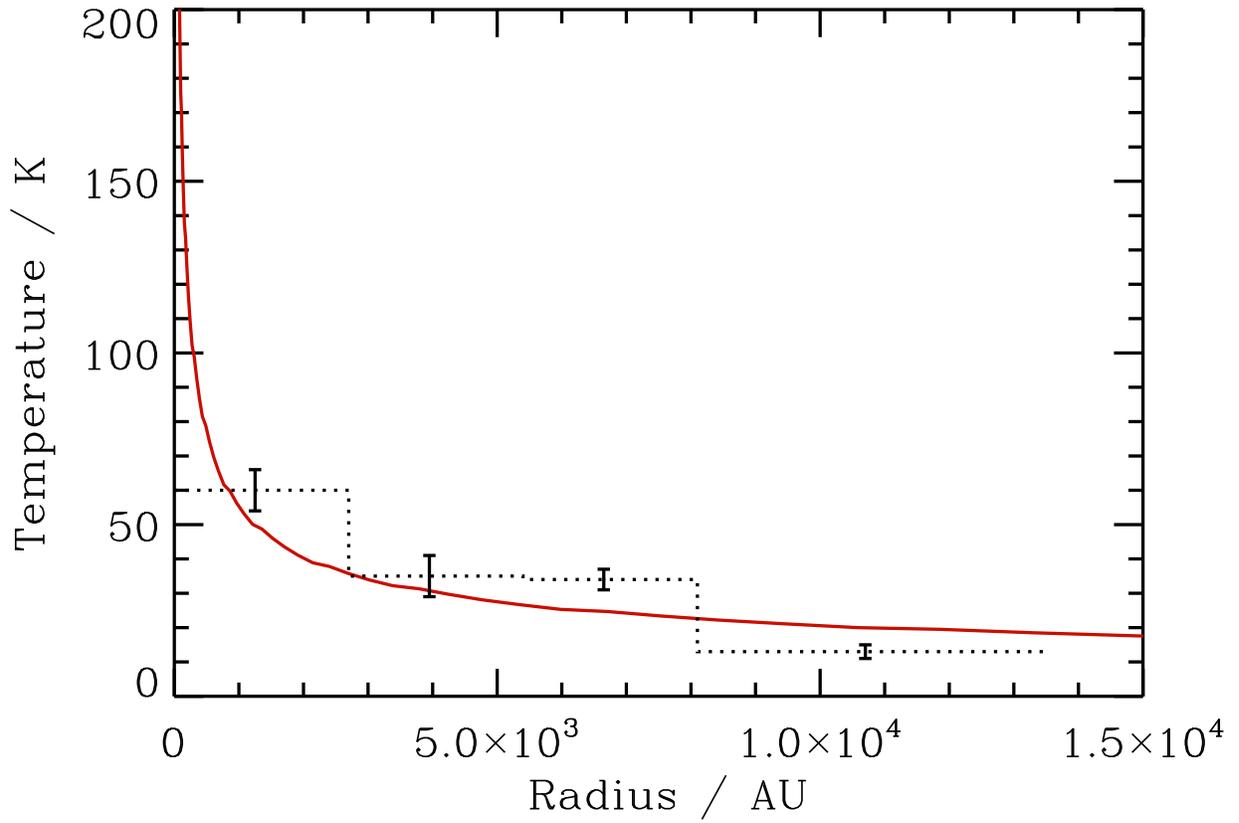}
\caption{The expected temperature structure in the NGC 7538 IRS 9 envelope \citep{vanderTak00} and the rotational temperatures derived from the CH$_3$OH data (dotted line). The arcsec-to-au conversion assumes a distance of 2.7 kpc. \label{fig:temp}}
\end{figure}

Figure \ref{fig:rot} shows the rotation diagrams for the four molecules in the four investigated annuli around the protostar. For all molecules the slope of the linear fit to the data points decreases at smaller scales, implying a higher excitation temperature closer to the central source (Table \ref{tab:cd}).  All molecules present rotational diagrams consistent with optically thin emission, i.e. the fluxes of strong and weak transitions can be fit by a single line.  The derived excitation temperatures span 13--39~K in the outer envelope, and are above 60~K at $R<1\arcsec$. While there is overlap between the different molecules in terms of upper energy levels, CH$_3$OH lines probe somewhat lower excitation levels on average, which may contribute to the lower temperatures derived for CH$_3$OH compared to the other molecules. CH$_3$OH is also known to be sub-thermally excited at low densities, and the outer envelope excitation temperature of 13~K should be taken as a lower limit of the kinetic temperature.
The CH$_3$OH excitation temperatures as a function of radius are shown in Fig. \ref{fig:temp} together with the expected kinetic temperature profile of the protostellar envelope from \citet{vanderTak00}, displaying the excellent fit.  

Figure \ref{fig:cd} displays the derived abundances of CH$_3$CCH, CH$_3$CN and H$_2$CS with respect to CH$_3$OH in different annuli. There is a small increase around 3$\arcsec$ or 8000~AU for H$_2$CS, followed by a constant abundance at smaller radii. The CH$_3$CN abundance clearly increase (with respect to CH$_3$OH) at the smallest scales, while the CH$_3$CCH abundance is constant at all radii (within the observational uncertainties).
 
\section{Discussion \label{sec:disc}}

\subsection{The Chemistry toward NGC 7538 IRS9}

NGC 7538 IRS9  contains a rich organic chemistry, including many molecules normally associated with hot cores. In this study we show the spatial distribution is highly molecule specific, and while IRS9 seems to contain a small hot core, most molecules, including CH$_3$CN, are present throughout the protostellar envelope; the observed CH$_3$CN distribution probably explains the need for a two-component fit (characterized by a hot (170~K) and luke-warm (80~K) excitation temperature) to CH$_3$CN lines toward the MYSO G11.92.0.61-MM1 \citep{Cyganowski11}. This implies that all detected complex molecules are at least in part zeroth and first generation ice chemistry products, with a potential contribution from second generation chemistry for molecules like CH$_3$CN that increase with respect to the ice product CH$_3$OH close to the protostar. In addition, molecules that are equally or more abundant outside of 3$\arcsec$ compared to the inner envelope and core are rightly classified as zeroth order molecules, since their formation must require very little heat. Examples are CH$_3$CCH, CH$_3$OH, CH$_3$CHO and CH$_2$CO (Fig. \ref{fig:cum}, Table \ref{tab:cd}). 

The observations thus suggest that the chemistry toward NGC 738 IRS9 develops from one dominated by hydrocarbons,  CH$_3$OH and unsaturated complex O-bering organics at large radii to a more and more saturated and N-rich organic chemistry at the core. This supports previous suggestions that spatial differentiation between O- and N-bearing complex molecules sometimes observed, trace differences in evolutionary stage between objects, but it is important to note that not all O-bearing molecules are the same. toward NGC 7538 IRS9 the chemical evolution is further observed to occur in two steps corresponding to $\sim$8000~AU and an unresolved central component. The latter can be attributed to a small hot core, while the former coincides with the temperature regime ($\sim$25~K), where complex ice chemistry (first generation molecule production) is expected to begin \citep{Garrod08}. 

The low S/N in the presented observation prohibits a quantitative analysis of many interesting molecules such as CH$_3$OCH$_3$. With ALMA this will become trivial for similar sources in the southern sky. In addition we will no longer we limited to the 1$\arcsec$ annuli analyzed here or the assumed spherical symmetry, since emission will readily be extracted on much smaller scale with higher S/N. The presented strategy is this likely to become a powerful tool to explore the chemical evolution around low- and high-mass protostars in the coming years.

\subsection{Model Comparison}

\begin{figure*}[htp]
\epsscale{1.0}
\plotone{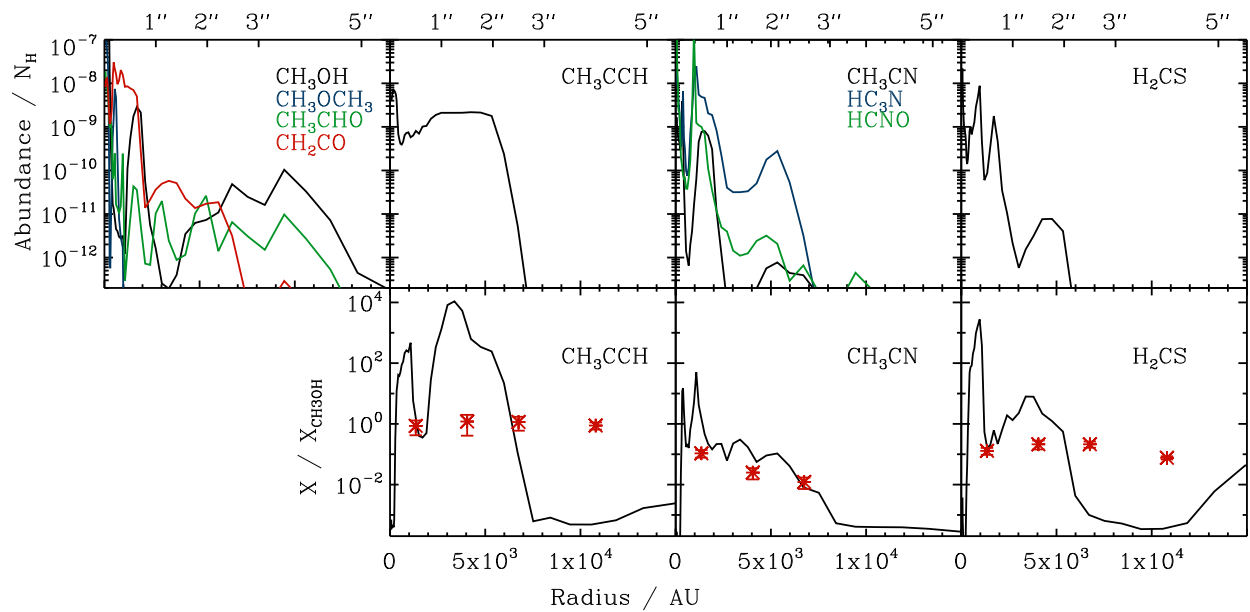}
\caption{Predicted organic abundances with respect to H$_2$ and CH$_3$OH toward NGC 7538 IRS 9. In the lower panel observations are overplotted in red. \label{fig:model}}
\end{figure*}

A crucial characteristic of the presented analysis strategy is its potential to test model predictions and thus benchmark both general model scenarios and formation and destruction pathways of specific molecules. Figure \ref{fig:model} shows an astrochemistry model from (Garrod 2013, ApJ accepted) mapped onto the NGC 7538 IRS9 temperature structure from \citet{vanderTak00}. The model output is a molecular abundance profile as a function of radius and temperature in the simulated protostellar envelope. Ignoring the model radii, the expected molecular abundance at each temperature point is shown as a function of distance from the central object using the  NGC 7538 IRS 9 temperature profile by \citet{vanderTak00}. The astrochemical model builds on \citep{Garrod08}, but incorporates several advances in our understanding of ice chemistry, especially a separation of ice surface and ice bulk chemistry, as well as new reaction pathways that were not previously considered.

Almost all complex organic molecule abundances display a large increase around 5000--7000 au, or 2--3$\arcsec$, in qualitative agreement with the observed increase in normalized emission around the same radius for many molecules. In the model, most molecules also present a second abundance jump a $r<1\arcsec$, in agreement with observations of e.g. CH$_3$CN and CH$_3$OH. The overall chemical structure thus seems consistent between observations and model.
 
A few molecules do appear to display inconsistent abundance patterns in the model compared to observed emission profiles, however. CH$_2$CO has an observed flat emission profile indicative of a zeroth generation molecule, while the model displays a clear increase with decreasing radius. The model also seems to miss cold formation pathways for CH$_3$CCH, HC$_3$N and H$_2$CS that can explain their observed envelope component. 

\begin{figure}[htp]
\epsscale{1.0}
\plotone{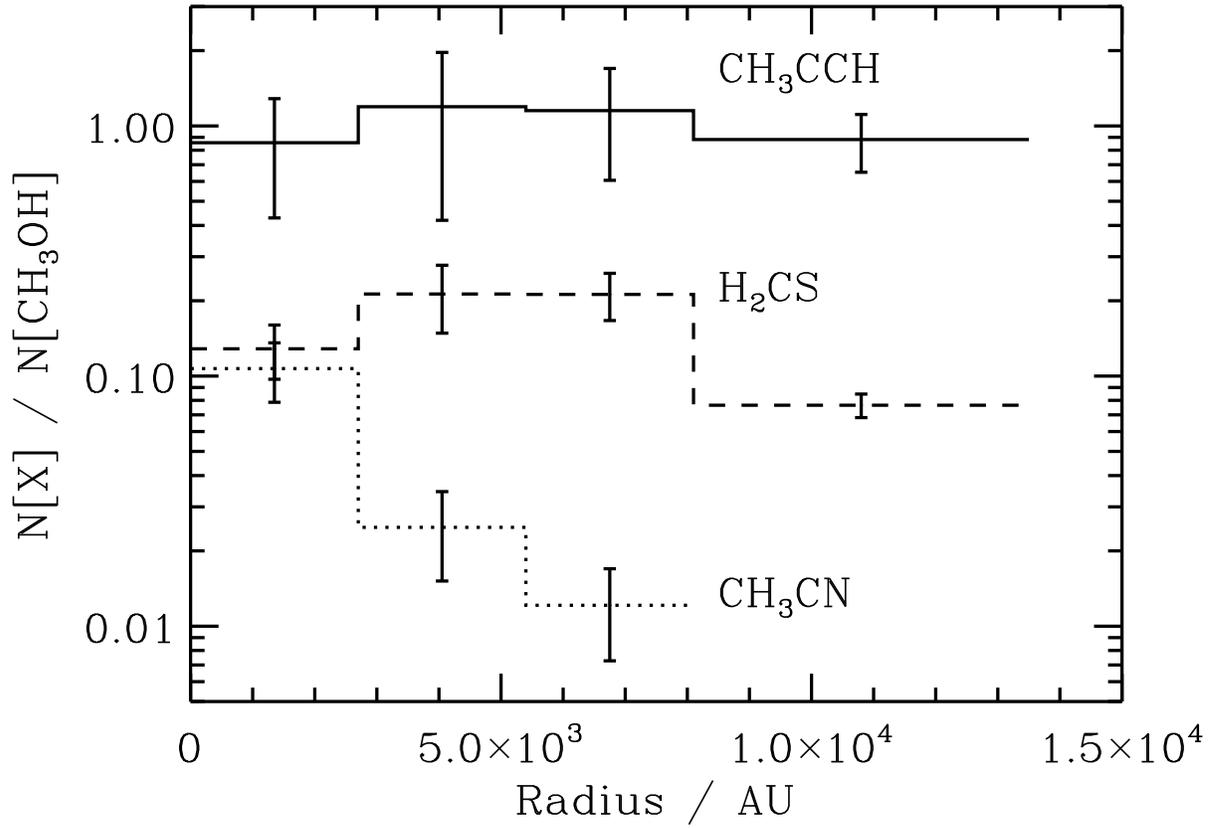}
\caption{The abundances with respect to CH$_3$OH toward NGC 7538 IRS 9 as a function of distance from the continuum peak. \label{fig:cd}}
\end{figure}

The lower panels of Fig. \ref{fig:model} display abundances of CH$_3$CN, CH$_3$CCH and H$_2$CS with respect to CH$_3$OH, which can be directly compared to Fig. \ref{fig:cd}. There is excellent agreement for CH$_3$CN/CH$_3$OH. There is also some qualitative agreement for the H$_2$CS/CH$_3$OH ratio; it increases at 3$\arcsec$ in both model and observations, but the increase is orders of magnitude larger in the model. The CH$_3$CCH/CH$_3$OH ratio is significantly different between model and observations, confirming that an important cold formation pathway is missing for CH$_3$CCH (and unsaturated hydrocarbons) in the model.

In summary, several of the observed trends are well reproduced, indicating that the main mode of complex molecule formation employed in the model, radical diffusion and reactions in ices, is the dominant formation pathway in this source. Some cold ($T<25$~K) formation pathways do seem to be missing however. This may be related to the recent detections of complex molecules in cold low-mass pre- and protostellar sources \citep{Oberg10a,Oberg11b,Bacmann12,Cernicharo12} and in infrared dark clouds \citep{Vasyunina11}, and should be the topic of further theoretical and experimental investigations. 

\section{Conclusions \label{sec:conc}}

Based on the observations, analysis and data-model comparison we draw the following conclusions:

\begin{enumerate}
\item NGC 7538 IRS9 presents a rich organic chemistry that is not spatially confined to a central unresolved component, a hot core, but rather extends through the luke-warm and cold protostellar envelope.
\item Based on the spatially resolved emission structures, the organic chemistry evolves from  hydrocarbons and unsaturated molecules in the outer envelop, to saturated molecules and N-bearing organics in the inner envelope.
\item The emission structures also reveal that in the envelope there is a dramatic change in the chemistry around 8000~AU (25~K), which corresponds to the onset of efficient ice chemistry in a recent protostellar chemstry model.
\item Determinations of excitation temperatures for CH$_3$OH and other molecules as a function of radius results in a steadily increasing temperature toward the core, consistent with a  heated protostellar envelope and in excellent agreement with an existing model of this object.
\item Quantification of emission fluxes, excitation temperatures and column densities at different radii is a useful tool to constrain the chemical evolution toward regularly shaped protostellar objects, and allows for direct comparison with chemical model predictions.
\end{enumerate}

\noindent The manuscript has benefitted from discussions with Ewine van Dishoeck. The SMA is a joint project between the Smithsonian Astrophysical Observatory and the Academia Sinica Institute of Astronomy and Astrophysics and is funded by the Smithsonian Institution and the Academia Sinica. RTG acknowledges support from the NASA Astrophysics Theory Program through grant NNX11AC38G.

\bibliographystyle{aa}
\bibliography{../../../BIB/mybib}


\begin{table*}[htp]
\begin{center}
\caption{Molecular line fluxes and line data$^{\rm a}$. \label{tab:flux}}
{\scriptsize
\begin{tabular}{l cccccccccc}
\hline \hline
Freq.&Log(A$_{ij}$)&E$_u$&d$_u$ $^{\rm b}$&FWHM&F$_{\rm r<5\arcsec}$&F$_{\rm r<3\arcsec}$&F$_{\rm r<2\arcsec}$&F$_{\rm r<1\arcsec}$\\
GHz& &K&&km s$^{-1}$&Jy km s$^{-1}$&Jy km s$^{-1}$&Jy km s$^{-1}$&Jy km s$^{-1}$\\
\hline
CH$_3$OH\\
241.700& -4.22&  47& 11&3.52$\pm$0.02&30.76$\pm$ 0.23&11.94$\pm$ 0.39&
 6.47$\pm$ 0.34& 3.40$\pm$ 0.22&\\
241.767& -4.24&  40& 11&3.61$\pm$0.01&65.03$\pm$ 0.21&15.86$\pm$ 0.39&
 8.17$\pm$ 0.35& 4.02$\pm$ 0.22&\\
241.791& -4.22&  34& 11&3.58$\pm$0.01&73.92$\pm$ 0.21&16.52$\pm$ 0.39&
 8.44$\pm$ 0.34& 4.06$\pm$ 0.22&\\
241.807& -4.66& 115& 22&3.57$\pm$0.40& 1.56$\pm$ 0.23& 0.76$\pm$ 0.39&
 0.89$\pm$ 0.35& 0.80$\pm$ 0.23&\\
241.833& -4.41&  84& 22&3.95$\pm$0.07&10.25$\pm$ 0.24& 7.24$\pm$ 0.44&
 4.30$\pm$ 0.38& 2.52$\pm$ 0.24&\\
241.852& -4.41&  97& 11&3.64$\pm$0.37& 1.75$\pm$ 0.23& 1.88$\pm$ 0.43&
 1.51$\pm$ 0.38& 1.07$\pm$ 0.24&\\
241.879& -4.22&  55& 11&3.87$\pm$0.03&25.09$\pm$ 0.22&10.73$\pm$ 0.41&
 5.74$\pm$ 0.36& 3.05$\pm$ 0.23&\\
241.888& -4.29&  72& 11&3.63$\pm$0.10& 6.58$\pm$ 0.23& 4.86$\pm$ 0.41&
 3.01$\pm$ 0.36& 1.85$\pm$ 0.23&\\
241.905& -4.30&  57& 11&3.72$\pm$0.02&31.84$\pm$ 0.24&15.31$\pm$ 0.41&
 8.38$\pm$ 0.36& 4.44$\pm$ 0.23&\\
CH$_3$CCH\\
239.179& -4.88& 201& 58&3.88$\pm$0.64& 1.11$\pm$ 0.24&$<$1.0&
$<$1.0&$<$0.66&\\
239.211& -4.00& 151& 16&3.31$\pm$0.11& 5.03$\pm$ 0.21& 2.85$\pm$ 0.37&
 1.58$\pm$ 0.32& 0.81$\pm$ 0.20&\\
239.234& -4.85& 115& 58&2.87$\pm$0.08& 5.42$\pm$ 0.20& 1.76$\pm$ 0.33&
 0.87$\pm$ 0.29& 0.42$\pm$ 0.18&\\
239.248& -4.84&  93& 58&3.05$\pm$0.05& 9.30$\pm$ 0.21& 3.06$\pm$ 0.34&
 1.62$\pm$ 0.30& 0.81$\pm$ 0.19&\\
239.252& -4.84&  86& 58&2.82$\pm$0.04&10.52$\pm$ 0.20& 3.48$\pm$ 0.32&
 1.85$\pm$ 0.29& 0.91$\pm$ 0.18&\\
CH$_3$CN\\
239.064& -2.97& 194& 54&8.77$\pm$1.75& 1.95$\pm$ 0.46& 2.55$\pm$ 0.90&
 2.32$\pm$ 0.80& 1.84$\pm$ 0.56&\\
239.096& -2.95& 144&108&6.02$\pm$0.27& 5.26$\pm$ 0.30& 5.29$\pm$ 0.57&
 3.72$\pm$ 0.49& 2.52$\pm$ 0.32&\\
239.120& -2.94& 108& 54&5.40$\pm$0.29& 4.09$\pm$ 0.28& 2.49$\pm$ 0.50&
 1.69$\pm$ 0.43& 1.08$\pm$ 0.28&\\
239.133& -2.93&  87& 54&4.63$\pm$0.24& 5.64$\pm$ 0.34& 4.13$\pm$ 0.47&
 2.98$\pm$ 0.40& 2.07$\pm$ 0.26&\\
239.138& -2.93&  80& 54&5.22$\pm$0.23& 7.33$\pm$ 0.37& 5.57$\pm$ 0.50&
 3.92$\pm$ 0.43& 2.68$\pm$ 0.27&\\
H$_2$CS\\
240.267& -3.69&  46& 15&3.37$\pm$0.04&11.59$\pm$ 0.20& 4.66$\pm$ 0.32&
 2.31$\pm$ 0.28& 1.01$\pm$ 0.18&\\
240.382& -3.72&  98& 15&3.57$\pm$0.17& 3.25$\pm$ 0.21& 2.05$\pm$ 0.34&
 1.09$\pm$ 0.29& 0.55$\pm$ 0.18&\\
240.393& -3.78& 164& 45&4.17$\pm$0.13& 5.64$\pm$ 0.23& 3.07$\pm$ 0.36&
 1.74$\pm$ 0.31& 0.89$\pm$ 0.20&\\
240.549& -3.72&  98& 15&3.80$\pm$0.18& 3.41$\pm$ 0.22& 1.48$\pm$ 0.34&
 0.82$\pm$ 0.30& 0.47$\pm$ 0.19&\\
CH$_3$CHO\\
242.106& -3.30&  83& 54&2.82$\pm$0.39& 1.23$\pm$ 0.22& $<$ 0.78&
 $<$ 0.51& $<$ 0.30&\\
242.118& -3.30&  83& 54&3.79$\pm$0.53& 1.43$\pm$ 0.26& $<$ 0.93&
 $<$ 0.60& $<$ 0.33&\\
 CH$_3$OCH$_3$\\
241.946& -3.78&  81&378&4.53$\pm$0.66& 1.42$\pm$ 0.31& 0.84$\pm$ 0.31&
 0.72$\pm$ 0.25& 0.45$\pm$ 0.17&\\
 HNCO\\
241.774& -3.71&  69& 23&3.03$\pm$0.26& 2.40$\pm$ 0.26& 1.85$\pm$ 0.42&
 1.18$\pm$ 0.50& 0.64$\pm$ 0.25&\\
 HC$_3$N\\
227.419& -3.03& 141& 51&5.37$\pm$0.09& 9.87$\pm$ 0.22& 6.56$\pm$ 0.30&
 4.26$\pm$ 0.20& 2.69$\pm$ 0.13&\\
 CH$_2$CO\\
240.186& -3.81&  87& 75&3.42$\pm$0.30& 2.01$\pm$ 0.23& $<$ 0.78&
 $<$ 0.54& $<$ 0.33&\\
\hline
\end{tabular}
\\$^{\rm a}$ From CDMS. $^{rm b}$ Degeneracy in the upper level.
}
\end{center}
\end{table*}

\begin{table*}[htp]
\begin{center}
\caption{Column densities in different annuli., \label{tab:cd}}
{\footnotesize
\begin{tabular}{l cccccccccc}
\hline \hline
Molecule&\multicolumn{2}{c}{5--3$\arcsec$}&\multicolumn{2}{c}{3--2$\arcsec$}&\multicolumn{2}{c}{2--1$\arcsec$}&\multicolumn{2}{c}{1--0$\arcsec$}\\
&N&T$_{\rm rot}$&N&T$_{\rm rot}$&N&T$_{\rm rot}$&N&T$_{\rm rot}$\\
\hline
\smallskip
CH$_3$OH&1.7$\pm$0.1e+15& 13[1]&8.3$\pm$0.7e+14& 34[2]&7.7e$\pm$0.8e+14& 35[3]&3.5$\pm$0.4e+15&60[6]\\
\smallskip

CH$_3$CCH&1.5$\pm$0.4e+15& 36[3]&9.6$\pm$4.6e+14&72[23]&9.2$\pm$6.0e+14&76[34]&3.8$\pm$1.0e+15&84[32]\\
\smallskip

CH$_3$CN&--&--&1.0$\pm$0.4e+13&60[15]&1.9$\pm$0.8e+13&125[65]&3.8$\pm$1.0e+14&281[218]\\
\smallskip

H$_2$CS&1.3$\pm$0.1e+14& 39[2]&1.7$\pm$0.4e+14&47[5]&1.6$\pm$0.5e+14&49[7]&4.4$\pm$1.1e+14&56[8]\\
\hline
\end{tabular}
}
\end{center}
\end{table*}

\newpage

\begin{appendix}

\section{IRAM 30m Observations}

\begin{figure}[htp]
\epsscale{1.0}
\plotone{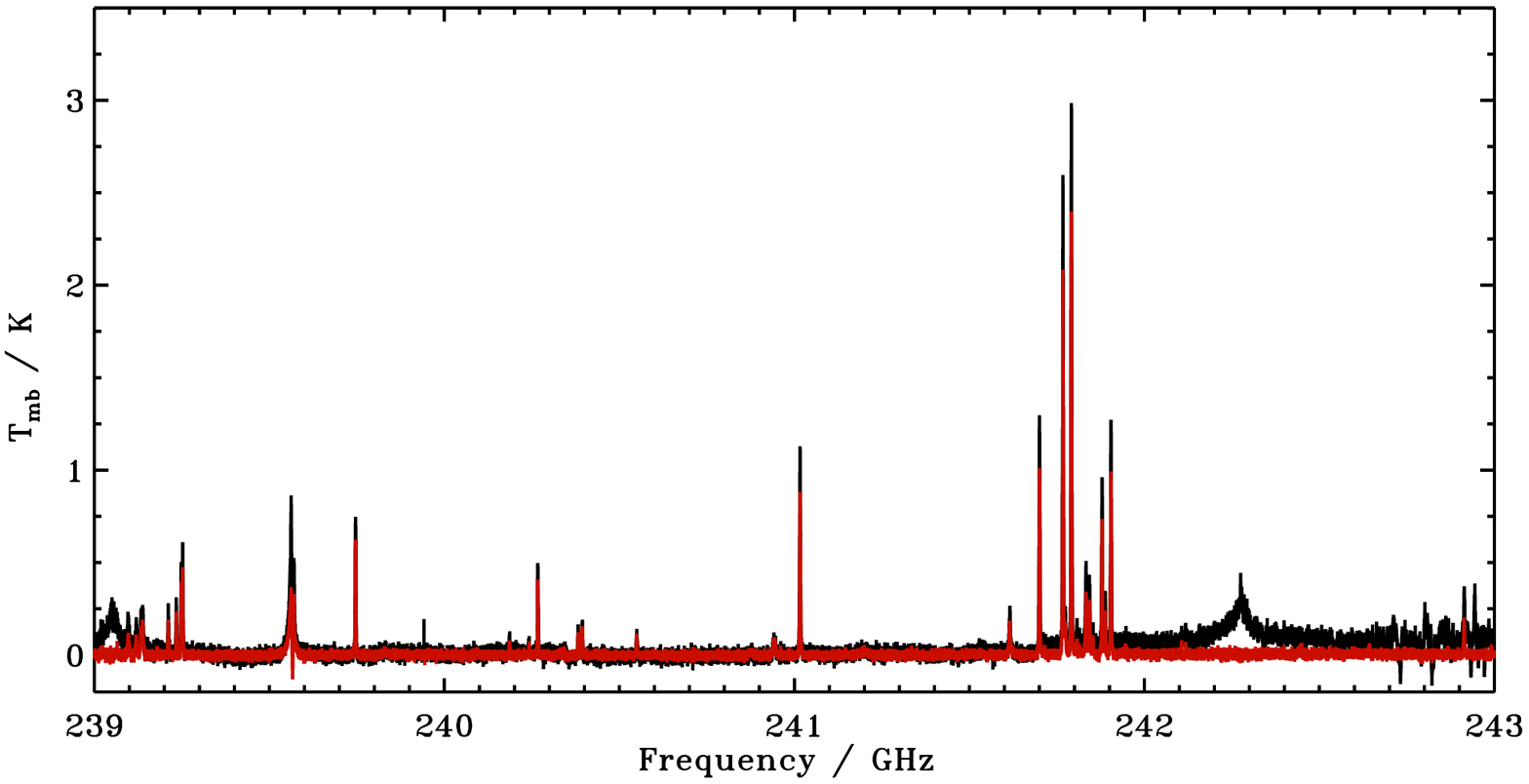}
\caption{IRAM 30m spectra toward NGC7538 IRS9, acquired in position switching (black) and wobbler mode. The position switching spectra is systematically 20\% more intense compared to the wobbler mode spectra. \label{fig:wob}}
\end{figure}

IRAM 30m spectra toward NGC 7538 IRS9 was acquired in both position switch and wobbler switch mode, with $\sim$30\% longer integration in the wobbler mode. Figure \ref{fig:wob} shows the 239--243~GHz part of the spectra, where most key lines for the purpose of this study are found. Above 242~GHz and below 239.2~GHz the position switch spectra display severe instabilities rendering it unusable. The remainder of the spectra display similar relative line intensities, but the lines in the wobbler spectra are consistently $\sim$20\% less intense on an absolute scale. Because the difference is homogenous (apart from the CO shadow line at 239.55~GHz) this likely due to calibration rather than emission in the wobbler off position. Thus the absolute calibration of the wobbler spectra is questionable, but the relative line intensities in the wobbler spectra should be accurate.

\begin{figure}[htp]
\epsscale{1.0}
\plotone{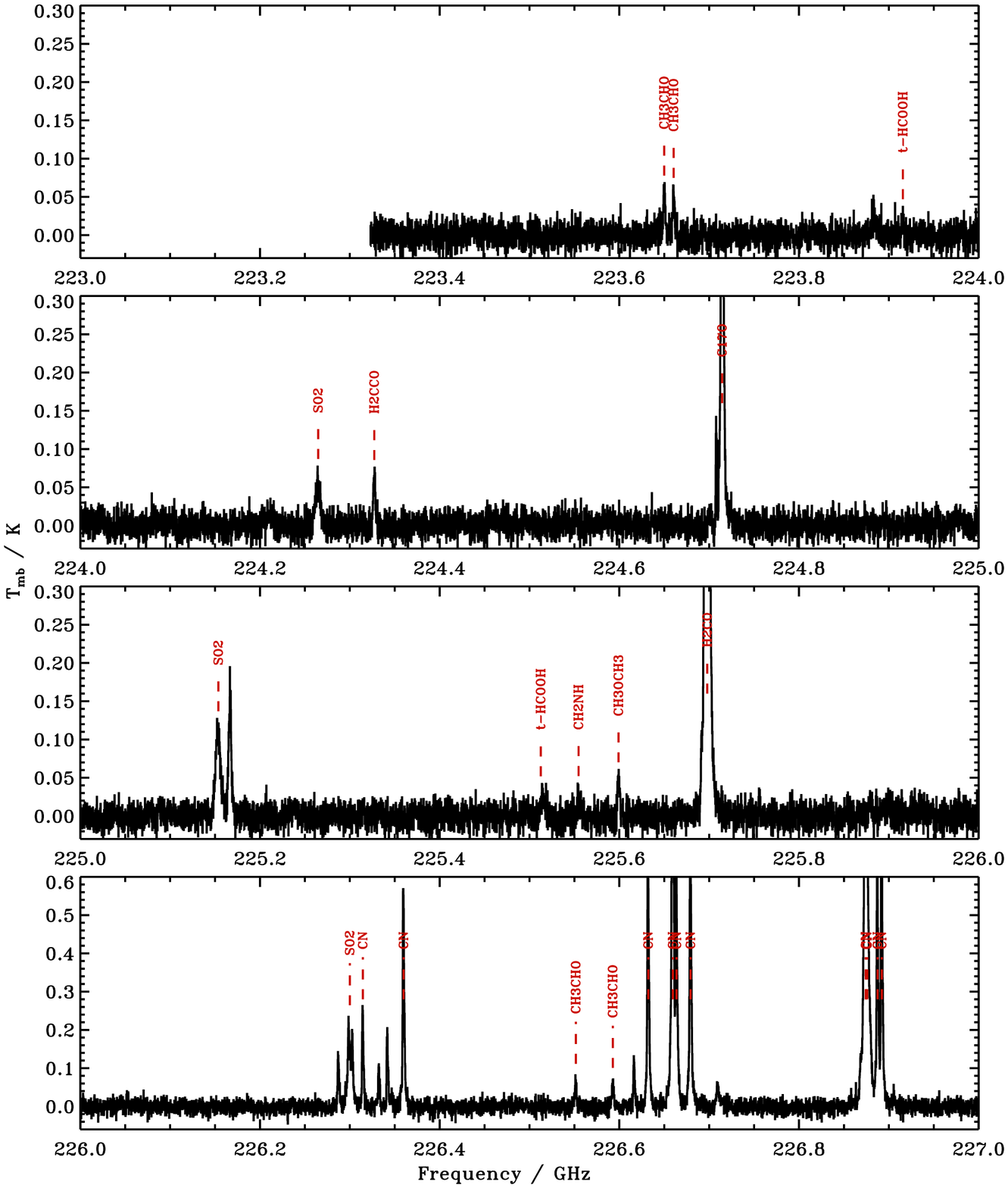}
\caption{IRAM 30m spectra toward NGC7538 IRS9 at 223--227 GHz displaying identified lines. \label{fig:lo}}
\end{figure}

\begin{figure}[htp]
\epsscale{1.0}
\plotone{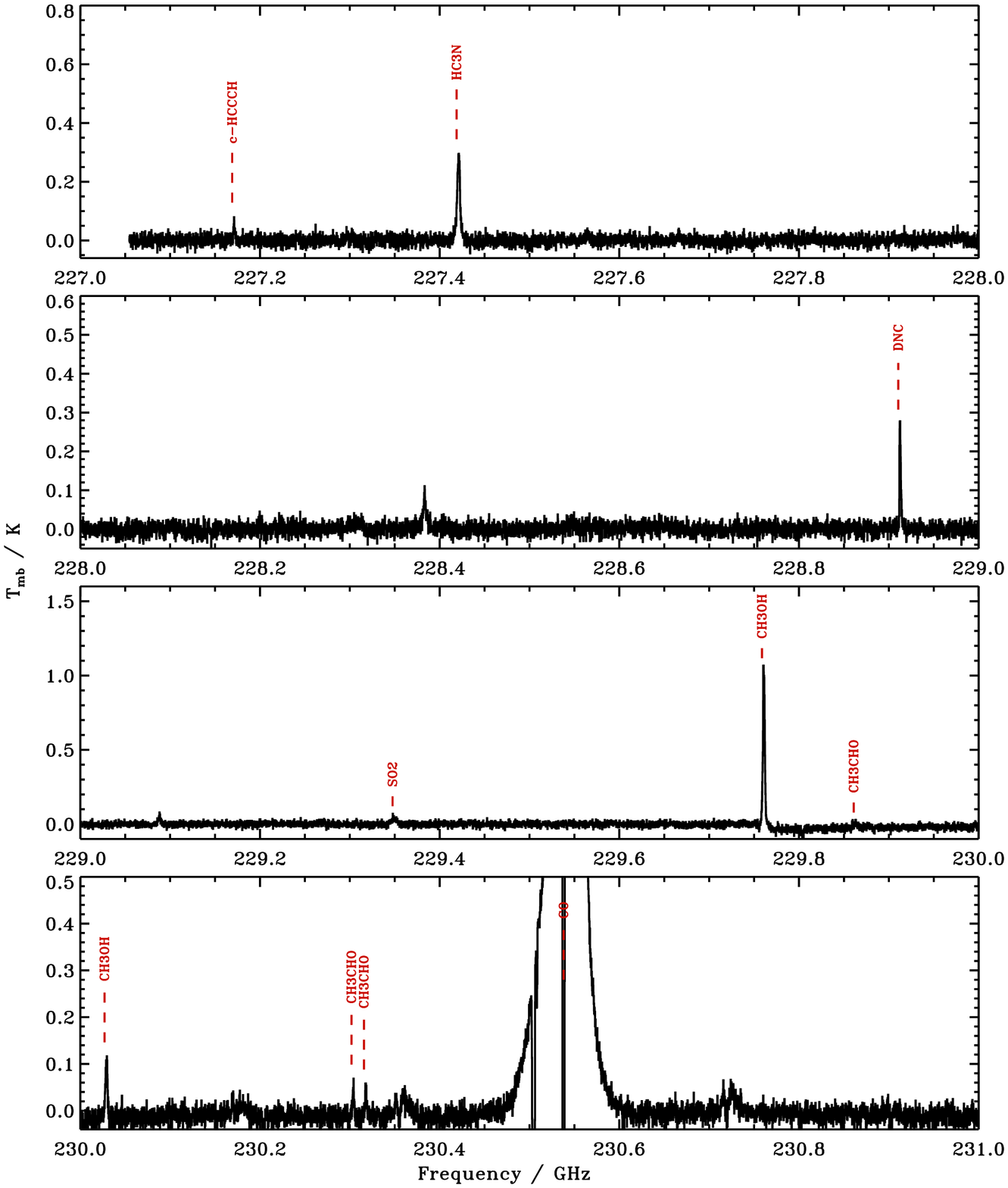}
\caption{IRAM 30m spectra toward NGC7538 IRS9 at 227--231 GHz displaying identified lines. \label{fig:li}}
\end{figure}

\begin{figure}[htp]
\epsscale{1.0}
\plotone{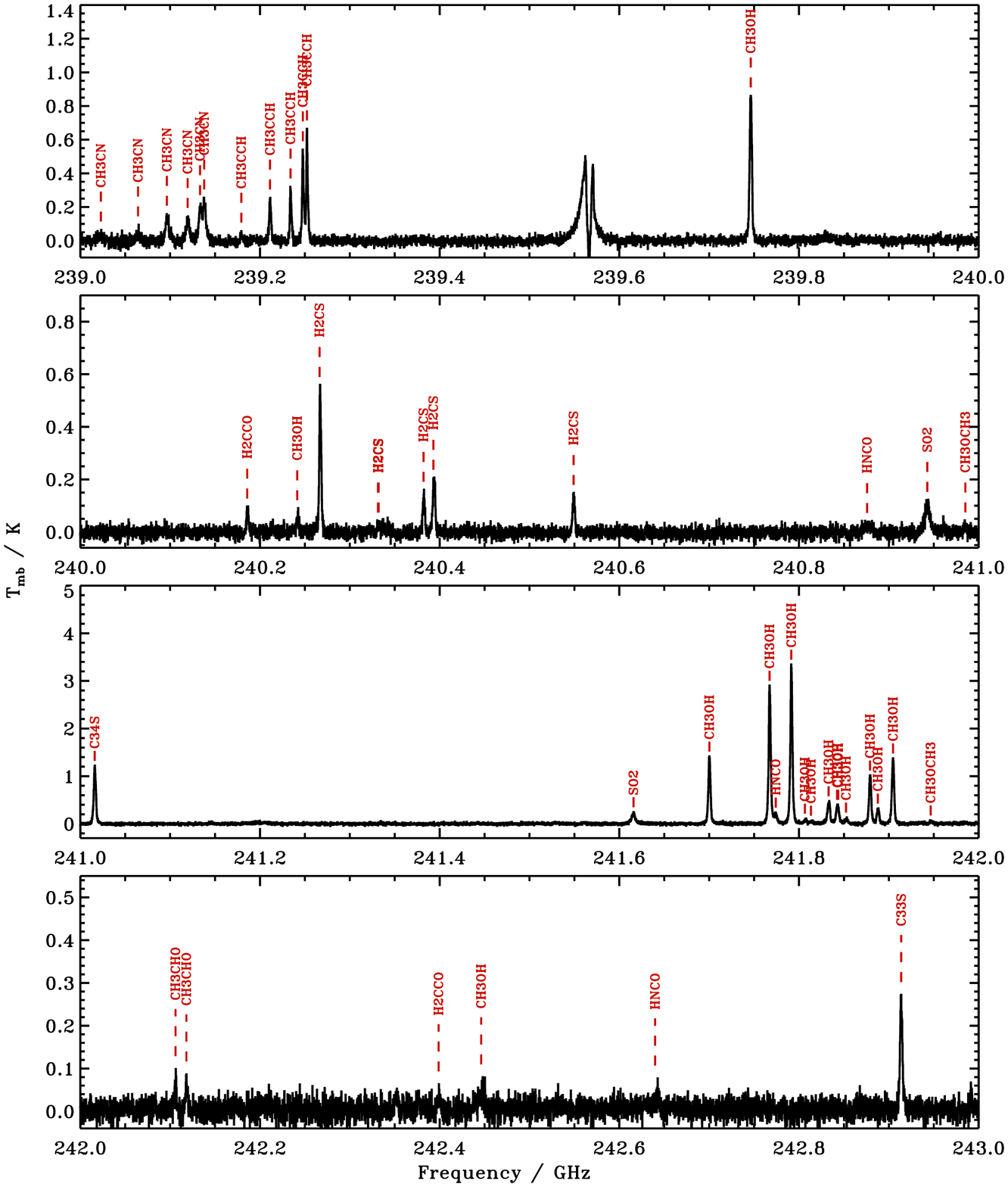}
\caption{IRAM 30m spectra toward NGC7538 IRS9 at 239--243 GHz displaying identified lines. \label{fig:ui}}
\end{figure}

\begin{figure}[htp]
\epsscale{1.0}
\plotone{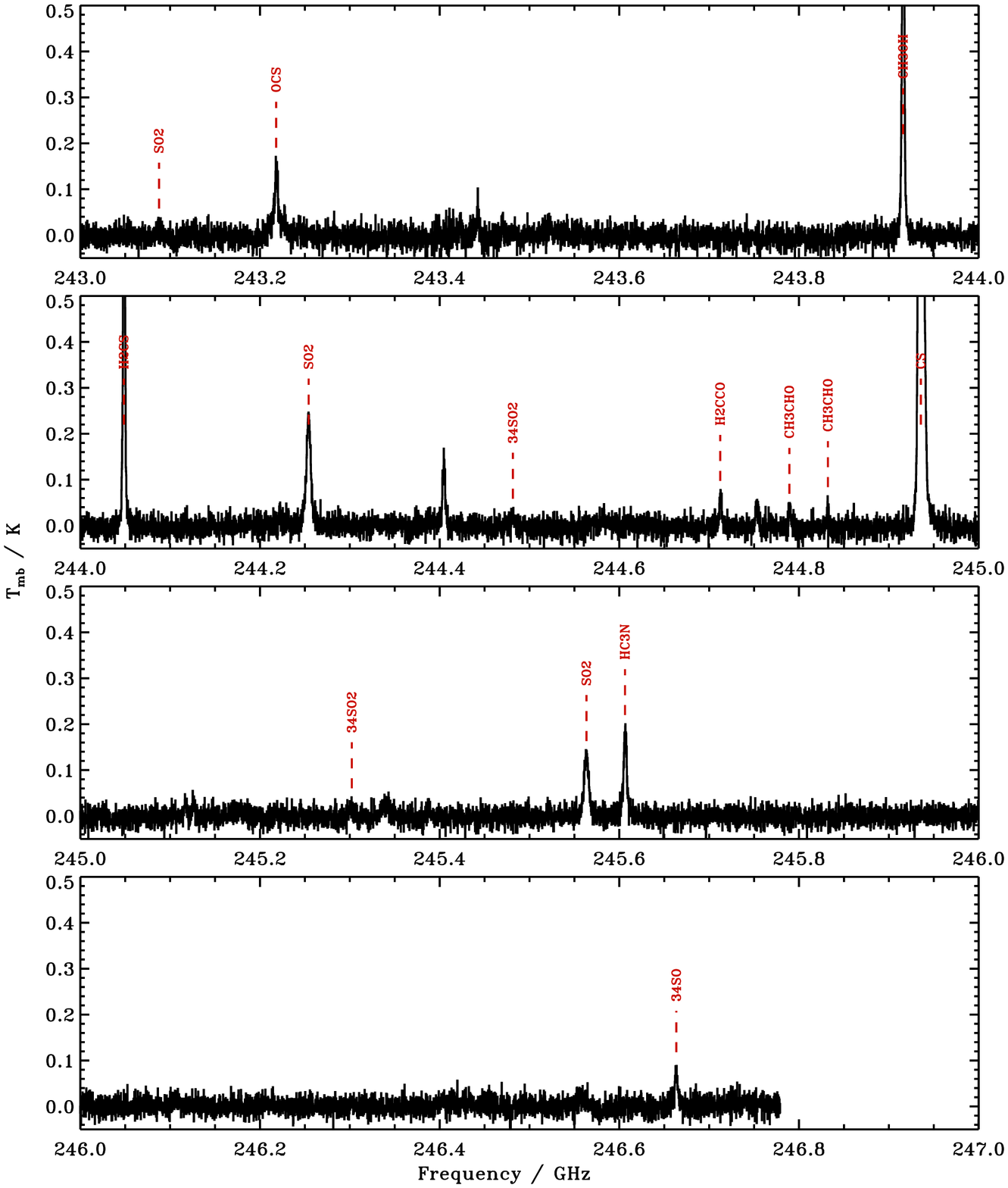}
\caption{IRAM 30m spectra toward NGC7538 IRS9 at 243--247 GHz displaying identified lines. \label{fig:uo}}
\end{figure}

Figures \ref{fig:lo}--\ref{fig:uo} present the complete IRAM 30m spectra acquired toward NGC7538 IRS9. Detections are marked and also listed in Tables \ref{det_l}--\ref{det_u}. Weak detections of species with only a few line in this range should be considered tentative. The line at 239.52~GHz is a ghost of the strong CO line in the lower sideband and the 244.44~GHz line is a ghost of the H$_2$CO line at 225.7 GHz. A few other minor features, e.g. at 228.38~GHz, could not be associated with any known spectral or ghost and remain unidentified.  

\begin{table}[ht]
\begin{center}
{\scriptsize
\begin{tabular}{l cccccccc}
\hline \hline
Species&freq [GHz]&E$_u$ [K]&d$_u$&A$_{\rm ul}$\\
\hline
CH$_3$CHO&223.6499&72.3&      50&-3.406\\
CH$_3$CHO&223.6604&72.2&      50&-3.406\\
t-HCOOH&223.9156&71.9&      21&-3.918\\
SO$_2$&224.2648&208&      41&-4.404\\
H$_2$CCO&224.3273&77.7&      69&-3.901\\
C$^{17}$O&224.7144&16.2&       5&-6.192\\
SO$_2$&225.1537&93.0&      27&-4.186\\
t-HCOOH&225.5126&88.1&      21&-3.932\\
CH$_2$NH&225.5546&10.8&       9&-4.031\\
CH$_3$OCH$_3$&225.5991&69.8&     200&-3.884\\
H$_2$CO&225.6978&33.4&      21&-3.557\\
SO$_2$&226.3001&119&      29&-3.972\\
CN&226.3146&16.3&       4&-5.004\\
CN&226.3599&16.3&       6&-4.794\\
CH$_3$CHO&226.5516&71.4&      50&-3.388\\
CH$_3$CHO&226.5927&71.3&      50&-3.388\\
CN&226.6322&16.3&       4&-4.371\\
CN&226.6596&16.3&       6&-4.024\\
CN&226.6637&16.3&       2&-4.072\\
CN&226.6793&16.3&       4&-4.278\\
CN&226.8742&16.3&       6&-4.017\\
CN&226.8748&16.3&       8&-3.942\\
CN&226.8759&16.3&       4&-4.066\\
CN&226.8874&16.3&       4&-4.564\\
CN&226.8922&16.3&       6&-4.742\\
c-HCCCH&227.1691&29.1&      27&-3.465\\
HC$_3$N&227.4189&142&      51&-3.029\\
DNC&228.9105&22.0&       7&-3.254\\
SO$_2$&229.3476&122&      23&-4.719\\
CH$_3$OH&229.7588&89.1&      17&-4.378\\
CH$_3$CHO&229.8609&71.3&      42&-4.390\\
CH$_3$OH&230.0271&39.8&       7&-4.828\\
CH$_3$CHO&230.3019&81.0&      50&-3.377\\
CH$_3$CHO&230.3158&81.1&      50&-3.377\\
CO&230.5380&16.6&       5&-6.161\\

\hline
\end{tabular}
}
\end{center}
\caption{Identified molecular lines at 223-231~GHz. Line information from CDMS and JPL spectral libraries.\label{det_l}}
\end{table}

\begin{table}[ht]
\begin{center}
{\scriptsize
\begin{tabular}{l cccccccc}
\hline \hline
Species&freq [GHz]&E$_u$ [K]&d$_u$&A$_{\rm ul}$\\
\hline
CH$_3$CN&239.0230&259&      58&-2.999\\
CH$_3$CN&239.0643&195&      50&-2.975\\
CH$_3$CN&239.0965&145&     108&-2.955\\
CH$_3$CN&239.1195&109&      50&-2.942\\
CH$_3$CN&239.1333&87.5&      54&-2.934\\
CH$_3$CN&239.1379&80.3&      50&-2.931\\
CH$_3$CCH&239.1793&202&      58&-4.879\\
CH$_3$CCH&239.2112&151&      16&-4.002\\
CH$_3$CCH&239.2340&115&      58&-4.851\\
CH$_3$CCH&239.2477&93.3&      58&-4.844\\
CH$_3$CCH&239.2523&86.1&      58&-4.842\\
CH$_3$OH&239.7463&49.1&      11&-4.247\\
H$_2$CCO&240.1858&88.0&      75&-3.810\\
CH$_3$OH&240.2415&82.5&      11&-4.841\\
H$_2$CS&240.2663&46.1&      15&-3.690\\
H$_2$CS&240.3315&256&      15&-3.861\\
H$_2$CS&240.3322&257&      15&-3.860\\
H$_2$CS&240.3821&98.8&      15&-3.725\\
H$_2$CS&240.3931&165&      45&-3.777\\
H$_2$CS&240.5491&98.8&      15&-3.724\\
HNCO&240.8758&113&      23&-3.720\\
SO$_2$&240.9428&163&      37&-4.153\\
CH$_3$OCH$_3$&240.9851&26.3&      88&-3.994\\
C$^{34}$S&241.0161&27.8&      11&-3.557\\
SO$_2$&241.6158&23.6&      11&-4.073\\
CH$_3$OH&241.7002&47.9&      11&-4.219\\
CH$_3$OH&241.7672&40.4&      11&-4.236\\
HNCO&241.7741&69.6&      23&-3.698\\
CH$_3$OH&241.7914&34.8&      11&-4.218\\
CH$_3$OH&241.8065&115&      11&-4.663\\
CH$_3$OH&241.8133&123&      11&-4.664\\
CH$_3$OH&241.8329&84.6&      11&-4.414\\
CH$_3$OH&241.8423&72.5&      11&-4.291\\
CH$_3$OH&241.8437&82.5&      11&-4.412\\
CH$_3$OH&241.8524&97.5&      11&-4.410\\
CH$_3$OH&241.8791&55.9&      11&-4.225\\
CH$_3$OH&241.8877&72.5&      11&-4.291\\
CH$_3$OH&241.9047&57.1&      11&-4.299\\
CH$_3$OCH$_3$&241.9465&81.1&     216&-3.781\\
CH$_3$CHO&242.1060&83.9&      54&-3.301\\
CH$_3$CHO&242.1182&83.8&      54&-3.301\\
H$_2$CCO&242.3989&193&      75&-3.823\\
CH$_3$OH&242.4462&249&      29&-4.640\\
HNCO&242.6399&113&      21&-3.698\\
C$^{33}$S&242.9138&28.1&       8&-3.584\\
SO$_2$&243.0877&53.1&      11&-4.989\\
OCS&243.2181&123&      41&-4.379\\
CH$_3$OH&243.9159&49.7&      11&-4.224\\
H$_2$CS&244.0485&60.0&      45&-3.677\\
SO$_2$&244.2542&93.9&      29&-3.786\\
34SO$_2$&244.4815&93.7&      29&-3.780\\
H$_2$CCO&244.7123&89.4&      75&-3.785\\
CH$_3$CHO&244.7893&83.1&      54&-3.286\\
CH$_3$CHO&244.8322&83.1&      54&-3.286\\
CS&244.9356&35.3&      11&-3.527\\
34SO$_2$&245.3023&40.7&      13&-3.991\\
SO$_2$&245.5634&72.7&      21&-3.924\\
HC$_3$N&245.6063&165&      55&-2.929\\
$^{34}$SO&246.6635&49.9&      11&-3.743\\
\hline
\end{tabular}
}
\end{center}
\caption{Identified molecular lines at 239--247~GHz. Line information from CDMS and JPL spectral libraries.\label{det_u}}
\end{table}

\end{appendix}

\end{document}